\documentclass[twocolumn]{aastex61}

\pdfoutput=1 
\usepackage{amsmath,amstext}
\usepackage[T1]{fontenc}
\usepackage{apjfonts} 
\usepackage{comment}
\usepackage[figure,figure*]{hypcap}
\usepackage{multirow}
\usepackage{amsmath}
\usepackage{nccmath}
\usepackage{rotating}
\usepackage{soul}
\usepackage{booktabs}
\setstcolor{red}
\usepackage{longtable}
\usepackage{dcolumn}

\newcommand{\HII}{\mbox{H\,{\sc ii}}}

\defcitealias{ortiz2017a}{OLK17}
\defcitealias{loinard2008}{LTM08}

\shorttitle{DYNAMO I: dynamical mass of Oph-S1}
\shortauthors{Ordóñez-Toro et al.}

\begin{document}

\title{Dynamical mass of the Ophiuchus intermediate-mass stellar system S1 with DYNAMO--VLBA}

\correspondingauthor{Jazmín Ordóñez-Toro}
\email{n.ordonez@irya.unam.mx}

\author[0000-0001-7776-498X]{Jazmín Ordóñez-Toro}
\affiliation{Instituto de Radioastronom\'{\i}a y Astrof\'{\i}sica, Universidad Nacional Aut\'onoma de M\'exico, Apartado Postal 72-3, Morelia 58089, M\'exico}
\affiliation{Departamento de Astronom\'{\i}a, Universidad de Guanajuato, Apartado Postal 144, 36000 Guanajuato, M\'exico}

\author[0000-0001-6010-6200]{Sergio A. Dzib}
\affiliation{Max-Planck-Institut f\"ur Radioastronomie, Auf dem H\"ugel 69, D-53121 Bonn, Germany}
\affiliation{IRAM, 300 rue de la piscine, F-38406 Saint Martin d'H\`eres, France}

\author[0000-0002-5635-3345]{Laurent Loinard}
\affiliation{Instituto de Radioastronom\'{\i}a y Astrof\'{\i}sica, Universidad Nacional Aut\'onoma de M\'exico, Apartado Postal 72-3, Morelia 58089, M\'exico}

\author[0000-0002-2863-676X]{Gisela Ortiz-León}
\affiliation{Instituto Nacional de Astrofísica, Óptica y Electrónica, Apartado Postal 51 y 216, 72000 Puebla, México }
\affiliation{Max Planck Institut f\"ur Radioastronomie, Auf dem H\"ugel 69, D-53121 Bonn, Germany}

\author[0000-0002-5365-1267]{Marina A. Kounkel}
\affiliation{Department of Astronomy, University of Michigan, 500 Church Street, Ann Arbor, MI 48105, USA}

\author{Josep M. Masqué }
\affiliation{Departamento de Astronom\'{\i}a, Universidad de Guanajuato, Apartado Postal 144, 36000 Guanajuato, M\'exico}

\author{S.-N. X. Medina }
\affiliation{Max Planck Institut f\"ur Radioastronomie, Auf dem H\"ugel 69, D-53121 Bonn, Germany}

\author[0000-0003-2271-9297]{Phillip A. B. Galli}
\affiliation{Núcleo de Astrofísica Teórica, Universidade Cidade de Sao Paulo, R. Galvao Bueno 868, Liberdade, 01506-000 Sao Paulo, SP, Brazil }

\author[0000-0001-9823-1445]{Trent J. Dupuy}
\affiliation{Institute for Astronomy, University of Edinburgh, Royal Observatory, Blackford Hill, Edinburgh, EH9 3HJ, UK}

\author[0000-0003-2737-5681]{Luis F. Rodr\'{\i}guez}
\affiliation{Instituto de Radioastronom\'{\i}a y Astrof\'{\i}sica, Universidad Nacional Aut\'onoma de M\'exico, Apartado Postal 72-3, Morelia 58089, M\'exico}

\author{Luis H. Quiroga-Nu\~nez}
\affiliation{Department of Aerospace, Physics and Space Sciences, Florida Institute of Technology A: 150 W University Blvd, Melbourne, 32901, FL, USA }

\begin{abstract}

We report dynamical mass measurements of the individual stars in the most luminous and massive stellar member of the nearby Ophiuchus star-forming region, the young tight binary system S1. We combine 28 archival datasets with seven recent, proprietary VLBA observations obtained as part of the \textit{Dynamical Masses of Young Stellar Multiple Systems with the VLBA} project (DYNAMO--VLBA), to constrain the astrometric and orbital parameters of the system, and recover high accuracy dynamical masses. The primary component, S1A,  is found to have a mass of 4.11$\pm$0.10~M$_\odot$, significantly less than the typical value, $\sim$~6~M$_\odot$ previously reported in the literature. We show that the spectral energy distribution of S1A can be reproduced by a reddened blackbody with a temperature between roughly 14,000~K and  17,000~K. According to evolutionary models, this temperature range corresponds to stellar masses between 4~M$_\odot$ and 6~M$_\odot$ so the SED is not a priori inconsistent with the dynamical mass of S1A. The luminosity of S1 derived from SED-fitting, however, is only consistent with models for stellar masses above 5~M$_\odot$. Thus, we cannot reconcile the evolutionary models with the dynamical mass measurement of S1A: the models consistent with the location of S1A in the HR diagram correspond to masses at least 25\% higher than the dynamical mass. For the secondary component, S1B, a mass of 0.831~$\pm$~0.014~M$_\odot $ is determined, consistent with a low-mass young star. While the radio flux of S1A remains roughly constant throughout the orbit, the flux of S1B is found to be higher near the apastron.
\end{abstract}

\section{Introduction}\label{intro}

The Ophiuchus molecular cloud complex is one of the nearest and best-studied sites of active star formation (see review by \citealt{wilking2008}). It has, indeed, played a crucial role in our overall understanding of star formation. Most of the ongoing star formation activity occurs in the Ophiuchus core (also known as Lynds 1688; L\,1688), located at 137.3 $\pm$ 1.2 pc \citep[][  \citetalias{ortiz2017a} hereafter]{ortiz2017a}. The most luminous and massive member of L\,1688, and of the entire Ophiuchus region, is the young Herbig-Be star Oph-S1 (also known as S1). Early radio observations of S1 showed that its radio emission originates mainly from two components: a thermal extended ($\sim20''$) component associated to a compact \HII\ region and a non-thermal compact source embedded  within the \HII\ region \citep{andre1988}. Very Long Baseline Interferometry (VLBI) observations later showed that the non-thermal component is associated with a magnetically active star; S1 was the first young stellar object directly detected with the VLBI technique \citep{andre1991}. These radio observations, as well as early infrared measurements by \citet{lada1984}, suggested that S1 was a young intermediate-mass star of spectral type B3 to B5, with a mass of about 6 M$_\odot$. 

While the properties of the compact radio emission in S1 are consistent with a magnetically active star, the detection of non-thermal radio emission in an intermediate-mass star is {\it a priori} unexpected. This is because the interiors of young intermediate-mass stars (3--8 M$_\odot$) are thought to be fully radiative while they move toward the main-sequence. Under these circumstances, the dynamo effect, which requires a rotating core surrounded by a convective envelope, cannot operate \citep{Stelzer2005, Stelzer2006, Hubrig2009}. Observational evidence, however, has shown that a significant fraction of Herbig Ae/Be stars and main sequence A/B stars show signs of magnetic activity \citep{Hubrig2009,Wade2011,hubrig2021}. A plausible explanation is that the magnetic activity is associated with a low-mass companion star rather than the intermediate-mass star itself \citep[e.g.,][]{Stelzer2005,Stelzer2006}. In the specific case of S1, lunar occultation observations have indeed revealed the existence of a lower mass companion with a separation around 20~mas \citep{richichi94, simon95}. More recently, Very Long Baseline Array (VLBA) high angular resolution observations obtained as part of the \textit{{Gould's Belt Distances Survey }}  \citepalias[GOBELINS; ][]{ortiz2017a} detected two stellar components in S1. From these data, \citetalias{ortiz2017a} obtained preliminary dynamical masses of $5.78\pm0.15$ and $1.18\pm0.10$ M$_\odot$ for the two stars in S1. While these VLBA observations confirmed the existence of a low-mass companion, they also show that both the low-mass companion and intermediate-mass star emit non-thermal radio emission. Thus, a radio emission mechanism immanent to the intermediate star remains necessary. Possible mechanisms include the existence of a fossil magnetic field, and the temporary apparition of a thin convective layer on the surface of the star \citep{palla1993,tout1995,skinner2004}.

Regardless of the exact emission mechanism, the binarity of S1 provides us with a rare opportunity to dynamically measure the mass of an Herbig Be star. In this paper, we aim at improving on the mass determination of \citetalias{ortiz2017a}. We present new VLBA observations of S1, obtained as part of the {\textit{Dynamical Masses of Young Stellar Multiple Systems with the VLBA} (DYNAMO - VLBA)\footnote{https://www3.mpifr-bonn.mpg.de/div/radiobinaries/intro.php}} project. To improve the constraints on the mass of the stars in S1, we combine these new observations with previous VLBA data, particularly those obtained as part of GOBELINS. In the rest of the paper, we will call the primary of the system S1A (this is the young Herbig Be star; \citetalias{ortiz2017a} and \citealt[][ \citetalias{loinard2008} hereafter]{loinard2008}) and the secondary component S1B.

\section{Observations and data reduction \label{sec:obs}}


The DYNAMO-VLBA observations (VLBA project code: BD215) used the same correlator setup as GOBELINS to record the radio continuum flux at $\lambda =$ 6.0~cm ($\nu =$ 5 GHz) using the VLBA of the National Radio Astronomy Observatory (NRAO).  S1 was observed six times, approximately every four months, from early 2018 to late 2019. During the sixth observation, the antennas on all sites were not pointing correctly for the first half of the observation time, and an additional observation was obtained four days later. Thus, S1 was observed a total of seven times as part of the DYNAMO-VLBA project. The observational details and the measured position are provided in Table \ref{Ts1_full}.

The total duration of each observation was three hours. The first and last 25 minutes were dedicated to observing about a dozen quasars distributed over the entire sky. The rest of the observation was spent on the target source and its gain calibrator (J1627-2427) in cycles of two minutes on-target and one minute on the calibrator. This specific gain calibrator was chosen because it is located in the core of L\,1688 at an angular distance of $10'$ from S1 \citep{dzib2013}. For all observations, the julian date reported in Table \ref{Ts1_full} corresponds to the mid-point of the three-hour observing run. 

As mentioned earlier, S1 was observed with the VLBA prior to the DYNAMO-VLBA project. Multi-epoch, phased-referenced observations were obtained as part of VLBA projects with codes BL128 (six epochs), BT093 (eight epochs) as well as the GOBELINS survey with code BL175 (14 epochs). These observations were first reported in \citetalias{loinard2008}, \citetalias{ortiz2017a}, and \citet{ortiz2018b} and we include them in our analysis. Thus, this paper considers a total of 35 VLBA observations of S1, distributed between 2005 to 2019. 

The initial data calibration followed standard procedures for phase-referenced VLBI observations and was carried out with the Astronomical Image Processing System (AIPS) software \citep{Greisen2003}. The steps have been described in detail in \citet{loinard2007}, \citet{dzib2010} and \citetalias{ortiz2017a}.


The phase calibrator used in the DYNAMO-VLBA and GOBELINS observations is located only $10'$ from the target source, but its flux density is only about $50$~mJy.  On the other hand, the observations of BL128 and BT093 used the stronger ($\sim1.0\,$ Jy) quasar J1625--2527 located $\sim1\rlap{.}^{\circ}1$ south S1. As a consequence, larger residual calibration errors are expected to affect the final images obtained from BL128 and BT093. S1A is a compact source with a flux density of about 6.0~mJy \citepalias{loinard2008,ortiz2017a} so, given the expected noise levels in our observations, it ought to be detected on individual baselines with a signal-to-noise ratio of order 5 
on integrations of two minutes. Self-calibration with a solution interval as short as two minutes can therefore be applied \citep[see][for further details]{Cornwell1989}. 

The self-calibration procedure used the following steps. First, we imaged the externally calibrated visibilities with a few clean components so the image contained only the strongest peak. Second, phase corrections on time intervals of 15 minutes are obtained using this image as a model. Then, the first and second steps are repeated two more times, decreasing the solution time interval to eight and two minutes, respectively. Final corrections in amplitude and phase are calculated, with time intervals of fifteen and eight minutes, using the final image of the previous step as the model. The final image is obtained using these corrections. 

\begin{figure*}[ht]
\includegraphics[scale=0.58]{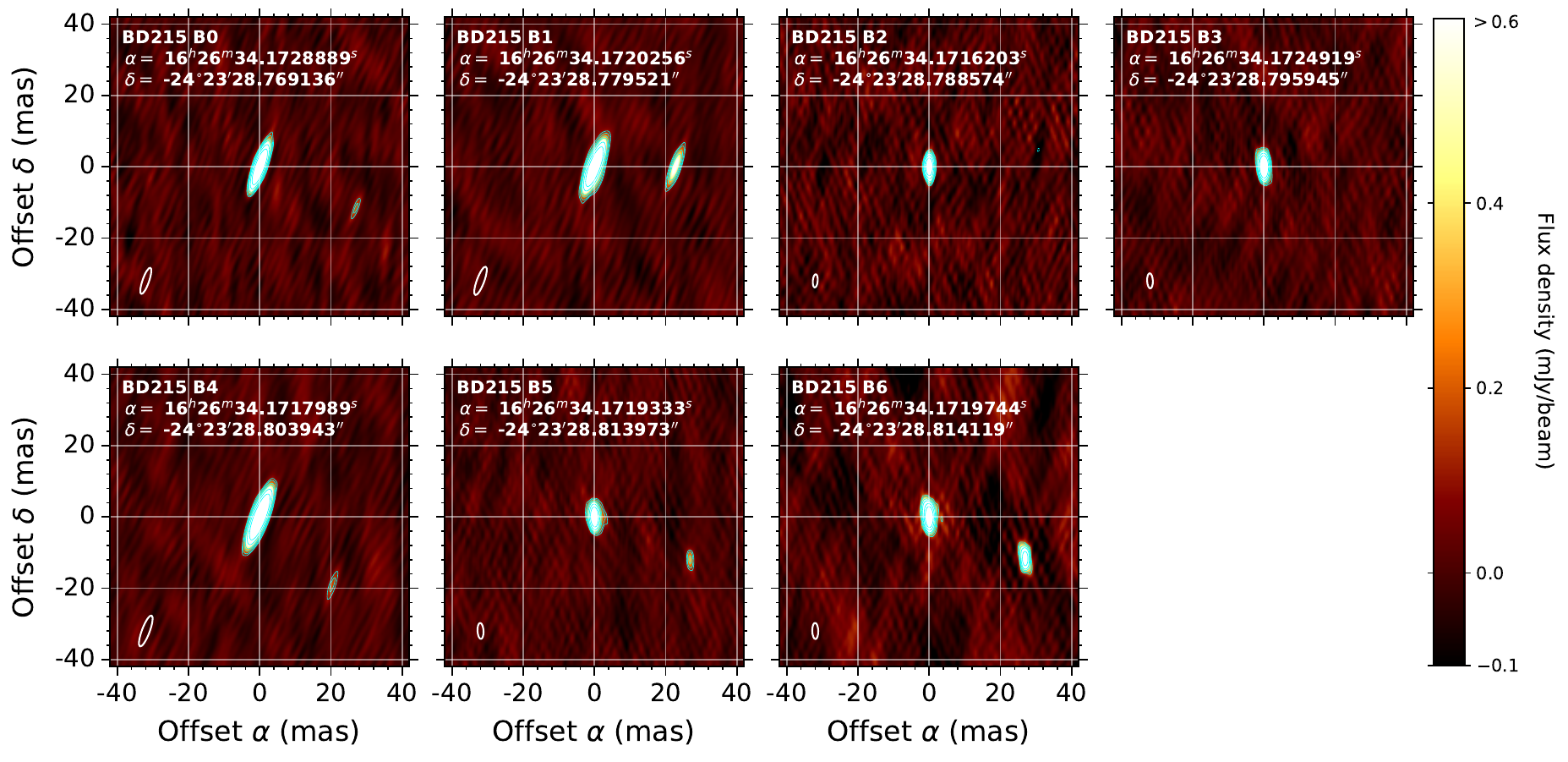}
\label{S1im}
\caption{Final radio images of S1 at 4.5 GHz corresponding to each epoch observed during the DYNAMO-VLBA project. Intensity background images are clipped to intensities between --0.1 to 0.6 mJy beam$^{-1}$.
Contour levels are at --4, 4, 6, 10, 15, 30, and 60 times the noise levels of the images, listed in Table~\ref{Ts1_full}. Images are centered on the measured position of S1A, the primary component of the S1 binary system as listed in Table~\ref{Ts1_full} and shown in the top-left corner of each plot. The white ellipse at the bottom-left of each plot represents the synthesized beam of the corresponding observed epoch.}
\end{figure*}

After self-calibration, the data were imaged using a pixel size of 100~$\,\mu$as and a natural weighting scheme (robust = 5 in AIPS). The rms noise level in the images (25--80~$\,\mu$Jy beam$^{-1}$, see Table~\ref{Ts1_full}) is consistent with the values expected theoretically. The fluxes and positions of the detected sources were measured using a two-dimensional Gaussian fitting procedure (task JMFIT in AIPS); the full list is presented in Table~\ref{Ts1_full}. Figure~\ref{S1im} shows the images of S1 as seen in the seven new epochs from the DYNAMO-VLBA project.

For the observations of projects BL128 and BT093 corresponding to the 14 older epochs, the measured positions were corrected to account for the most recent measured position of the gain calibrator used for these observations, J1625--2527. The positional offset was estimated by \citetalias{ortiz2017a} to be $\Delta \alpha = -23\,\mu$as and $\Delta \delta = -1.2\,$mas. These  values were added to the measured positions of S1A and S1B in all epochs corresponding to projects BL128 and BT093, and are already taken into account in the values listed in Table~\ref{Ts1_full}. 
\begin{table*}[ht]
		\caption{S1A and S1B measured positions and flux densities from the 35 VLBA observations.}
		\label{Ts1_full}
	\footnotesize
	\begin{center}
 \setlength{\tabcolsep}{2.8pt}
 \renewcommand{\arraystretch}{1.05}
	\begin{tabular}{lcccccccccc} \hline\hline
        &        &          &\multicolumn{3}{c}{S1A}&\multicolumn{3}{c}{S1B} & \\ \cmidrule(lr{.75em}){4-6}\cmidrule(lr{.75em}){7-9}
Project  &Date UT&	Date	& $\alpha$(J2000.0) &  $\delta$ (J2000.0) &$S_\nu$ &$\alpha$(J2000.0) &  $\delta$ (J2000.0) &$S_\nu$\tablenotemark{a} &$\sigma_{\rm noise}$	 \\  
name &(yyyy.mm.dd)& Julian Day& $16^{h}26^{m}$ [$^{s}$]&$-24^{\circ}$23$'$ [$''$]&(mJy) &$16^{h}26^{m}$ [$^{s}$]&$-24^{\circ}$23$'$ [$''$]&(mJy) &{\footnotesize ($\mu$Jy\,bm$^{-1}$)} \\ \hline
BL128 GA  & 2005.06.24 & 2453545.731  & 34.17392792(28) & 28.428288(10) & $11.05\pm0.14$&34.1727518(144) & 28.447731(383) & $~~0.97\pm0.27$&62\\
BL128 GB & 2005.09.15 & 2453628.504  & 34.17367442(61) & 28.433862(21) & $~~5.86\pm0.16$& \nodata & \nodata & $<0.23$ & 75\\
BL128 GC & 2005.12.17 & 2453722.248  & 34.17434762(96)  & 28.442703(36) &$~~5.94\pm0.18$& \nodata & \nodata & $<0.17$ &57\\
BL128 GD & 2006.03.15 & 2453810.007  & 34.17463122(73) & 28.452589(29) & $~~5.92\pm0.18$& \nodata & \nodata & $<0.20$ &68\\
BL128 GE & 2006.06.03 & 2453889.789  & 34.17400238(61) & 28.457124(20) & $~~5.77\pm0.17$& 34.17268902(578) & 28.453578(178) & $~~1.01\pm0.21$&75\\
BL128 GF & 2006.08.22 & 2453969.571 & 34.17328928(51) & 28.463940(18) & $~~6.90\pm0.15$& \nodata & \nodata & $<0.23$ &75\\
BT093 BA & 2007.06.05 & 2454256.784 & 34.17387398(51) & 28.480866(18) & $~~6.42\pm0.13$&34.17222738(509) & 28.498965(275) & $~~0.62\pm0.14$&70\\
BT093 BB & 2007.06.09 & 2454260.773 & 34.17402998(67) & 28.478336(24) & $~~6.15\pm0.18$&\nodata & \nodata & $<0.23$ &75\\
BT093 BC & 2007.06.13 & 2454264.762  & 34.17380163(30) & 28.481044(12) & $~~9.39\pm0.13$&\nodata & \nodata & $<0.20$ &67\\
BT093 BD & 2007.06.17 & 2454268.751 & 34.17386498(54) & 28.480123(17) & $~~7.38\pm0.15$&\nodata & \nodata & $<0.21$ &69\\
BT093 BE & 2007.06.21  & 2454272.741  & 34.17374613(40) & 28.481819(15) & $~~8.30\pm0.15$&34.17203044(111) & 28.499143(43) & $~~2.50\pm0.08$&79\\
BT093 BF & 2007.06.25 & 2454276.730 & 34.17377508(36) & 28.483839(13) & $10.36\pm0.15$&34.17205001(165) & 28.500992(66) & $<0.21$ &70\\
BT093 BG & 2007.06.29 & 2454280.718 & 34.17368822(56) & 28.482562(22) & $~~7.88\pm0.15$&34.17193311(538) & 28.49900(22) & $~~1.02\pm0.17$&74\\
BT093 BH & 2007.07.03 & 2454284.708  & 34.17364237(47) & 28.482137(17) & $~~7.82\pm0.15$&\nodata & \nodata & $<0.23$ &75\\
BL175 BN & 2012.12.16 & 2456278.239  & 34.17329746(93) & 28.629756(29) & $~~1.80\pm0.07$&\nodata & \nodata & $<0.08$ &25\\
BL175 E1 & 2013.09.02 & 2456537.533  & 34.17210801(57) & 28.648987(19) & $~~3.38\pm0.06$&\nodata & \nodata & $<0.08$ &27\\
BL175 G0 & 2014.03.01  & 2456718.038  & 34.17336345(48) & 28.661664(17) & $~~3.13\pm0.05$&\nodata & \nodata & $<0.08$ &28\\
BL175 CR & 2014.10.07 & 2456937.933  & 34.17244123(90) & 28.677675(30) & $~~3.41\pm0.09$&\nodata & \nodata & $<0.12$ &41\\
BL175 EX  & 2015.02.27 & 2457081.043 & 34.17321421(58) & 28.692174(21) & $~~3.28\pm0.06$&\nodata & \nodata & $<0.11$ &35\\
BL175 GW  & 2015.10.04 & 2457300.443  & 34.17207617(62) & 28.702069(22) & $~~2.79\pm0.05$&\nodata & \nodata & $<0.08$ &25\\
BL175 F5 & 2016.03.26 & 2457473.996  & 34.17309283(23) & 28.718105(8) & $~~6.44\pm0.04$&34.17127710(467) & 28.733943(184) & $~~0.27\pm0.02$&30\\
BL175 F8  & 2016.04.28 & 2457506.878 & 34.17291981(39) & 28.720696(13) & $~~4.41\pm0.05$&34.17100551(704) & 28.733448(334) & $~~0.17\pm0.03$&30\\
BL175 IJ & 2016.09.09 & 2457640.513 & 34.17202055(95) & 28.730246(30) & $~~4.56\pm0.08$&\nodata & \nodata & $<0.09$ &30\\
BL175 IN  & 2016.09.16 & 2457647.520   & 34.17201100(30) & 28.730546(11) & $~~5.65\pm0.05$&34.17029175(616) & 28.732533(285) & $~~0.20\pm0.03$&25\\
BL175 JE & 2017.03.18 & 2457831.019  & 34.17259913(35) & 28.745253(13) & $~~4.58\pm0.05$&\nodata & \nodata & $<0.09$ &28\\
BL175 JF & 2017.03.25 & 2457838.063  & 34.17259863(25) & 28.745577(9) & $~~6.95\pm0.05$&\nodata & \nodata & $<0.08$ &27\\
BL175 K8  & 2017.09.03 & 2457999.557   & 34.17175982(45) & 28.753704(17) & $~~3.95\pm0.05$&\nodata & \nodata & $<0.20$ &66\\
BL175 KB & 2017.09.15 & 2458012.492  & 34.17183287(140) & 28.755294(74) & $~~5.58\pm0.10$&\nodata & \nodata & $<0.09$ &30\\
BD215 B0 & 2018.02.15 & 2458165.043 & 34.17288885(44) & 28.769136(14) & $~~6.73\pm0.05$&34.1709178(126) & 28.780844(399) & $~~0.20\pm0.03$&30\\
BD215 B1 & 2018.06.27 & 2458296.683  & 34.17202556(52) & 28.779521(16) & $~~6.45\pm0.05$&34.17036273(380) & 28.779616(22) & $~~0.73\pm0.03$&26\\
BD215 B2 & 2018.10.30 & 2458422.339  & 34.17162025(30) & 28.788574(10) & $~~8.30\pm0.08$&\nodata & \nodata & $<0.11$ &38\\
BD215 B3 & 2019.03.01 & 2458544.026  & 34.17249189(21) & 28.795945(7) & $~~8.85\pm0.06$&\nodata & \nodata & $<0.09$ &30\\
BD215 B4 & 2019.07.08 & 2458672.653  & 34.17179886(47) & 28.803943(14) & $~~7.42\pm0.05$&34.1703085(132) & 28.823218(465) & $~~0.19\pm0.03$&26\\
BD215 B5 & 2019.11.10 & 2458798.313  & 34.17193330(32) & 28.813973(11) & $~~5.56\pm0.05$&34.16996475(474) & 28.826151(186) & $~~0.38\pm0.05$&29\\
BD215 B6 & 2019.11.14 & 2458802.299   & 34.17197439(31) & 28.814119(11) & $~~9.79\pm0.09$&34.17000425(112) & 28.825751(39) & $~~2.75\pm0.09$&48\\
\hline
	\end{tabular}  
 \end{center}
 \tablenotetext{a}{Upper-limits are three times the noise level in the image. }
\end{table*}

\section{Astrometric fitting procedure}

The displacements of S1A and S1B on the celestial sphere are described by the combination of their common trigonometric parallax $\pi$, the uniform proper motion of their center of mass in right ascension and declination, $\mu_\alpha $ and $ \mu_\delta $, and their orbital motion. 
For the primary component, the complete equations describing the motions are:

\begin{align}
\alpha(t) & =\alpha_0 + \mu_\alpha t + \pi f_\alpha(t) + a_1 Q_\alpha(t),\label{eqn:pm1}\\
\delta(t) & =\delta_0 + \mu_\delta t + \pi f_\delta(t) + a_1 Q_\delta(t),\label{eqn:pm2}
\end{align}
\noindent where $ \alpha_0 $ and $ \delta_0 $ are the coordinates of the center of mass of the system at the chosen reference epoch, $ f_\alpha $ and $ f_\delta $ are the projections of the parallactic ellipse \citep{Seidelmann1992}, and $a_1$ is the semi-major axis of the orbit of the primary around the center of mass. $Q_\alpha(t)$ and $Q_\delta (t)$ represent the projections of the orbital motions and are functions of the orbital elements:

\begin{align}
Q_\alpha(t) & =r\left[\cos{\left(\theta+\omega\right)}\sin{\Omega}-\sin{\left(\theta+\omega\right)}\cos{\Omega}\cos{i}\right]/\cos{\delta},\label{eqn:Q1}\\
Q_\delta(t) & =r\left[\sin{\left(\theta+\omega\right)}\sin{\Omega}\cos{i}+\cos{\left(\theta+\omega\right)}\cos{\Omega}\right],\label{eqn:Q2}
\end{align}
\noindent where $\theta$ is the true anomaly and $r$ is the radius from the center of mass along a Keplerian orbit. The inclination angle is $i$, the angle of the line of nodes $\Omega$, and the angle from the ascending node to periastron $\omega$. For the secondary component, the semi major axis $a_2$ is used instead of $a_1$ in equations (\ref{eqn:pm1}) and (\ref{eqn:pm2}) and $\theta$ is rotated $180^\circ$ in equations (\ref{eqn:Q1}) and (\ref{eqn:Q2}) (see e.g.\ \citealt{kounkel2017} for details). 


To solve the previous equations we translated the procedure described by \citet{kounkel2017} from IDL to python\footnote{{The full routine is available at \url{https://github.com/mkounkel/astrometric_binaries/blob/master/astrometry_binary_python3.ipynb}}}. This routine simultaneously fits equations (\ref{eqn:pm1}) to (\ref{eqn:Q2}) to the data by the method of least squares using MPFIT. It is able to simultaneously process absolute positions of both stars from VLBA astrometry, as well as relative positions of the secondary with respect to the primary from optical data if they exist. 
The free orbital parameters in this model are the period $P$, the semimajor axis of primary $a_1$, the eccentricity $e$, the inclination $i$, the angle of the line of nodes $\Omega$, the time of periastron passage $T$, the angle from node to periastron $\omega$, and the mass ratio $m_2/m_1$. The total mass of the system is derived from Kepler's third law and the distance -- obtained from the very same fit through the trigonometric parallax.

This routine is sensitive to the initial guesses of several non-linear orbital parameters, namely $P$, $e$, $\omega$, and $\Omega$; with poorly chosen initial guesses the fitter may not converge, getting stuck in a local minimum instead. The remaining parameters are not affected by this, and can always be scaled. Thus, to fully probe the parameter space, we performed the fit 1000 different times with different initial guesses. The routine then iterates on these guesses, enabling optimal convergence for several dozens of these sets of iterations, which are then evaluated using $\chi^2$. These sets of iterations with $\chi^2<1.1\chi_{\rm best}^2$ are selected (where $\chi_{\rm best}^2$ is the minimum chi squared of the entire set. The uncertainties are evaluated in two ways: from examining the intrinsic scatter from all of the valid solutions, and from calculating a weighted average error of all of the errors returned by MPFIT, weighted by $\chi^2$. Typically, both methods produce comparable distributions. \\

%
\section{Results}
The S1 system was detected in all seven new VLBA observations (Figure~\ref{S1im}). S1A was detected in all these epochs, with a mean flux density of 7.59 mJy; S1B was detected five times with a mean flux density of 0.85 mJy. 
In the ancillary observations, thanks to the improvement obtained by applying self-calibration, we have detected S1B in more epochs than previously reported.  In addition, we found that a previous detection of S1B (BL128 GE from 2006 June 3) was spurious: the emission peak at that position in fact corresponded to a prominent side-lobe. We do detect S1B in the self-calibrated image of the same epoch but at a different position. Therefore, the measured S1B position on 2006 June 03  by \citetalias{ortiz2017a} is discarded in our analysis and we use, instead, the newly measured position from the self-calibrated image. S1A was detected at all 35 epochs -- a 100\% detection rate was also achieved in the publications reporting on older observations \citepalias[][\citealt{ortiz2018b}]{loinard2008,ortiz2017a}. On the other hand, the companion, S1B, is detected in 14 epochs, of which only four detections were reported by \citetalias{ortiz2017a}. All the measured positions and flux densities for S1A and S1B are listed in Table~\ref{Ts1_full}. 

\subsection{Astrometry}

\begin{figure}[!hbt]
\includegraphics[scale=0.35]{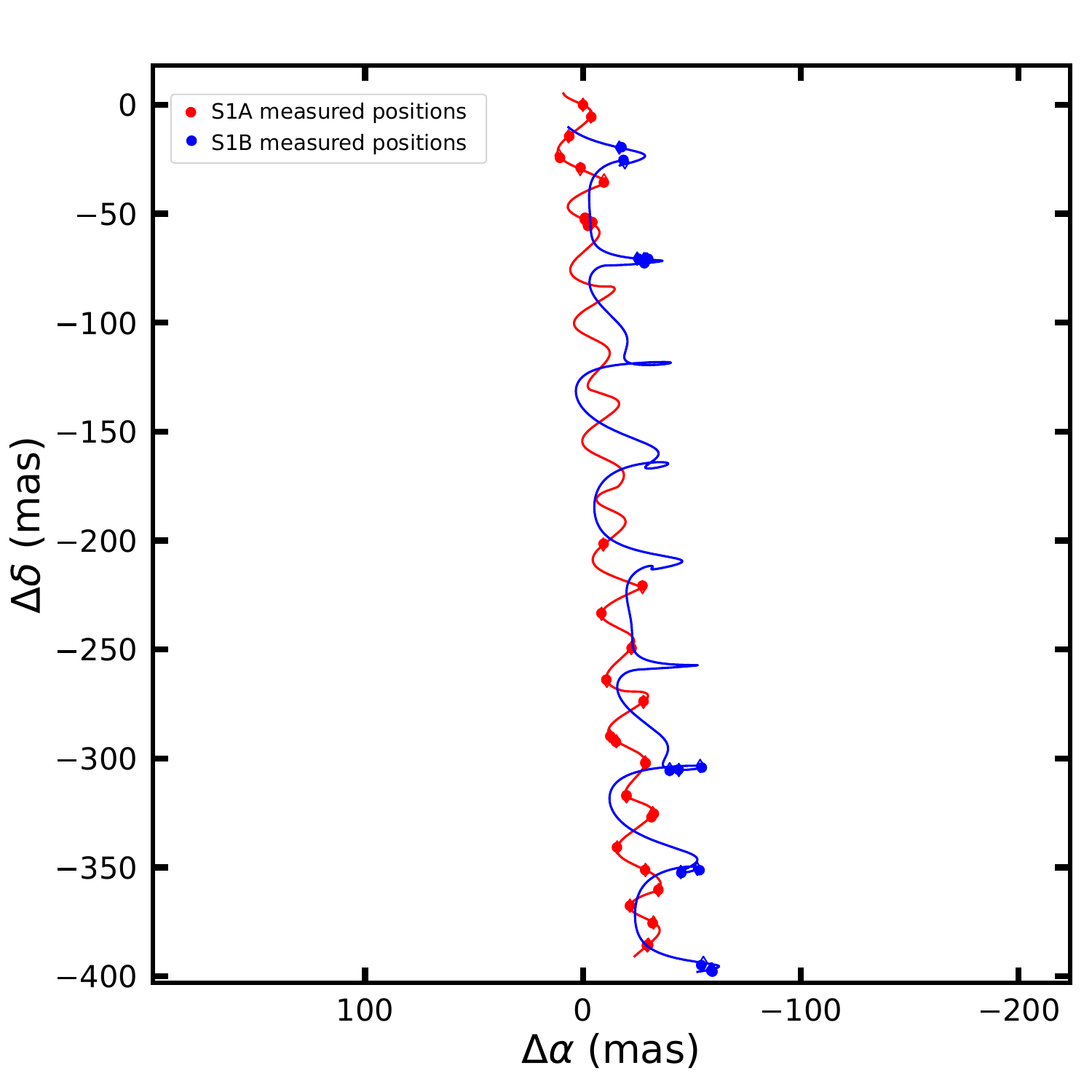}\\
\label{S1sky}
\caption{Measured positions of S1A (red dots) and S1B (blue dots) shown as offsets from the position of S1A in the first detected epoch used in the present paper (2005 June 24; see Table~\ref{Ts1_full}). 
The red and blue curves show the best fits, described in the text, to the positions of S1A and S1B, respectively.}
\end{figure}

\begin{figure}[htp!]
\includegraphics[scale=0.49
,trim=10 20 0 20,clip]{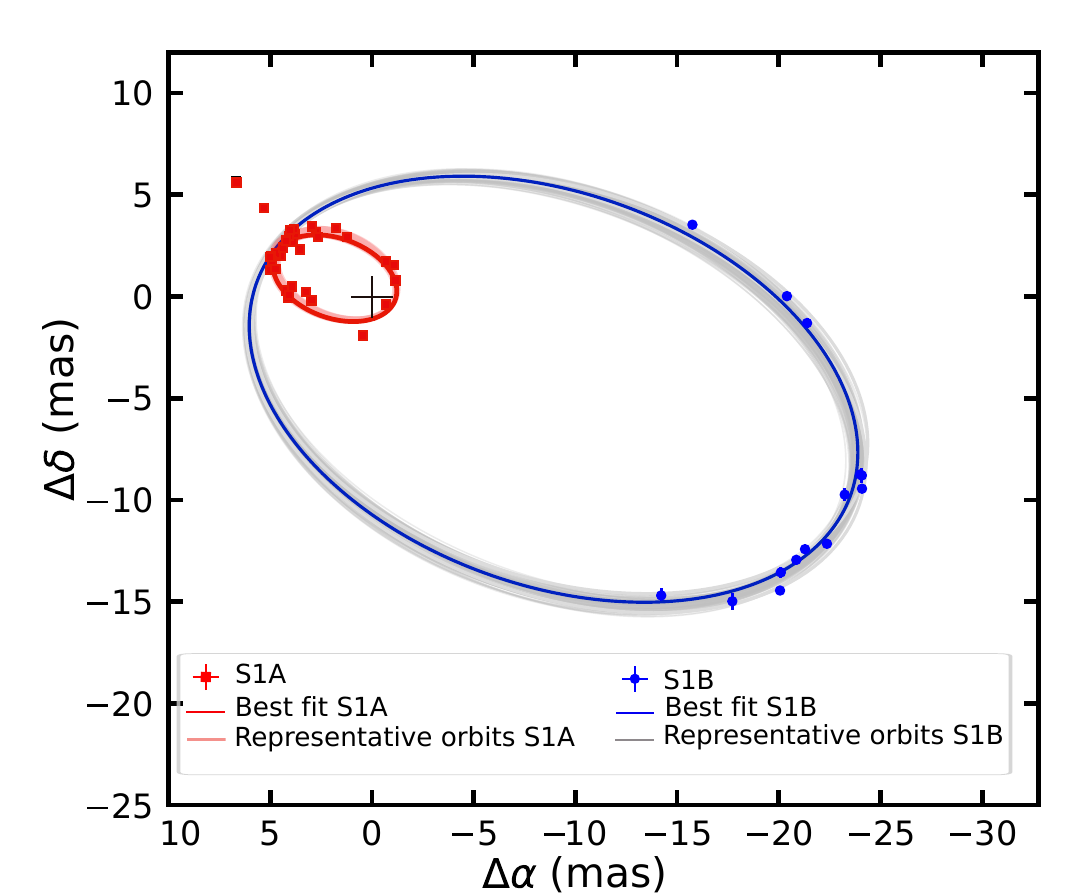}
\includegraphics[scale=0.33
,trim=0 0 0 0,clip]{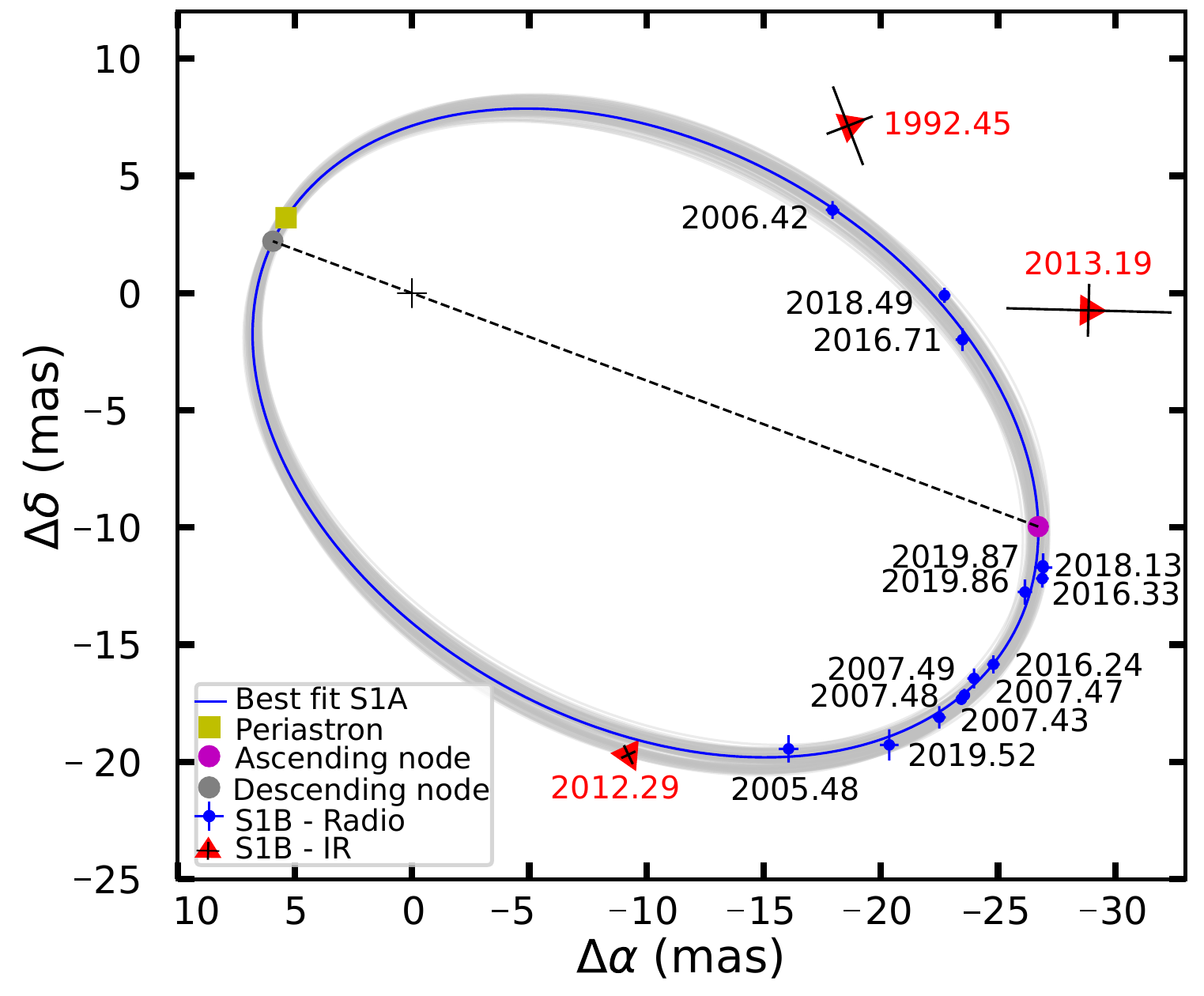}\\
\label{S1or}
\caption{Orbital solution of S1. {\it Top-panel:} Stellar orbits relative to the central of mass (black cross). The orbits from best fit for S1A (red) and S1B (blue), are accompanied by a subset of 100 orbit pairs (light red for S1A, and grey for S1B) to show the range of solutions allowed by the errors in our fit. The subset of orbits shown corresponds to the best 100 of 5,000 orbits generated by randomly obtaining orbital parameters from normal distributions with mean values equal to the values obtained from our best fit and with standard deviations equal to their errors.  Measured positions of S1A (red squares) and S1B (blue circles) after correction of parallax and proper motions are also shown. Position errors are on the order of 0.1 mas, barely seen at the plot scale. 
{\it Bottom-panel: } Stellar relative positions and orbital fit model of S1. The blue dots indicate the relative positions of S1B with respect to S1A, and the errorbars consider the position errors of both components which are added in quadrature. The dashed black line traces the line of nodes from the model, and the black cross indicates the position of the primary. The measured positions of S1B obtained from infrared observations reported in \citet{richichi94} and \citet{cheetham2015} are also plotted, as red triangles. These infrared positions are listed in Table~\ref{tab:IR}. }
\end{figure}

The final measured positions of S1A and S1B were used as inputs to our astrometric fitting procedure.  The resulting parameters from our best fit are listed in Table~\ref{T_A0} and are plotted in Figures~\ref{S1sky} (total sky motion)  and \ref{S1or} (orbit motion). The total and individual masses are derived from Kepler's third law, taking into account the obtained values for the period, semi-major axis, and the mass ratio from our model fit. As a result, we obtain masses for the primary and secondary components of 4.11$\pm$0.10~M$_\odot$ and 0.83$\pm0.01$, respectively. It is important to note that these mass estimates are somewhat at odds with those reported by \citetalias{ortiz2017a}: 5.78$\pm $0.15 M$_\odot$ for the primary component and 1.18$\pm$0.10 M$_\odot $ for the secondary component. This discrepancy arises from two main factors. Firstly, we have significantly increased the number of S1B detections, yielding more robust and accurate measurements. Secondly, we have corrected the position of S1B at epoch 2006 June 3rd (JD 2453889.789). As mentioned earlier, the detection of S1B reported by \citetalias{ortiz2017a} turned out to be spurious, and the new position is significantly different from the previous one. It is worth mentioning that \citetalias{ortiz2017a} based their fits on only four detections of S1B, so the inclusion of a spurious position had an especially large detrimental effect. The addition of seven new, high-quality data, obtained as part of DYNAMO--VLBA was particularly important to constrain the orbital fit. From our study, we conclude that S1A is indeed an intermediate-mass star, but less massive than previously suggested.\\

We also fitted the data with the MCMC Orbitize! package \citep{Blunt2017} to assess the accuracy of the orbital parameters obtained from the MPFIT fitting. This algorithm explores the multidimensional parameter space, effectively sampling the posterior distribution of the orbital parameters. The results obtained demonstrate remarkable consistency with the values derived from the MPFIT fit, confirming the reliability of our orbital parameter estimates. A detailed description of the MCMC analysis and its results is presented in Appendix \ref{app:1}.

\begin{table}[!hbt]
	\renewcommand{\arraystretch}{1.1}
		\caption{\footnotesize{Parameters obtained for the best-fit astrometric model.}}
		\label{T_A0}
	\centering
	\small
	\begin{tabular}{cccc} \hline \hline
		&Parameter& Value & Units 	  \\ 
		\hline
		\multicolumn{2}{l}{Astrometric parameters} &&\\
		& $\alpha_{2016.0, {\rm centre}}$ & 16:26:34.1723678(23)& hh:mm:ss\\
		& $\delta_{2016.0, {\rm centre}}$ & --24:23:28.712919(22) & $^\circ:\,':\,''$\\
		&$\Delta\alpha_{\rm S1A}$\tablenotemark{a}	& $3.69\pm0.04$&mas\\
		&$\Delta\delta_{\rm S1A}$\tablenotemark{a}& $3.25\pm0.02$&mas\\
		&$\Delta\alpha_{\rm S1B}$\tablenotemark{a}& $-17.99\pm0.04$&mas\\
		&$\Delta\delta_{\rm S1B}$\tablenotemark{a}&$-15.87\pm0.02$ &mas\\
		&$\mu_{\alpha}$  & $-2.536\pm0.002$ & mas yr$^{-1}$\\	
		&$\mu_{\delta}$  & $-26.853\pm0.001$ & mas yr$^{-1}$\\
		&$\pi$           & $7.28\pm0.02$ & mas\\
		&d & $137.2\pm0.4$ & pc \\
		\multicolumn{2}{l}{Orbital parameters}&&\\
		&$a_1$ & $0.418\pm0.002$ & AU \\
		&$a_2$ & $2.043\pm0.018$ & AU \\
	    &P & $1.734\pm0.001$ & years\\
	    &$T_0$ & $2457163.69\pm0.65$ & Julian date \\
	    &$e$ & $0.65\pm0.01$ & \\
	    &$\Omega$ & $249.6\pm6.3$ & degrees \\
	    &$i$ & $25.4\pm1.9$ & degrees \\
	    &$\omega$ & $168.4\pm6.4$ &degrees \\
	    \multicolumn{2}{l}{Dynamical masses}&&\\
	    &Total mass & $4.942\pm0.113$ & M$_\odot$ \\
	    &Mass 1 & $4.112\pm0.099$ & M$_\odot$ \\
	    &Mass 2 & $0.831\pm0.014$ & M$_\odot$ \\
	    \hline
	\end{tabular}  
	\tablenotetext{a}{$\Delta\alpha$ and $\Delta\delta$ values are the orbital offset relative to the mass center at the epoch 2016.0 of components S1A and S1B.  }
\end{table} 

\subsection{Flux variations}

\begin{figure}[!hbt]
\includegraphics[scale=0.71]{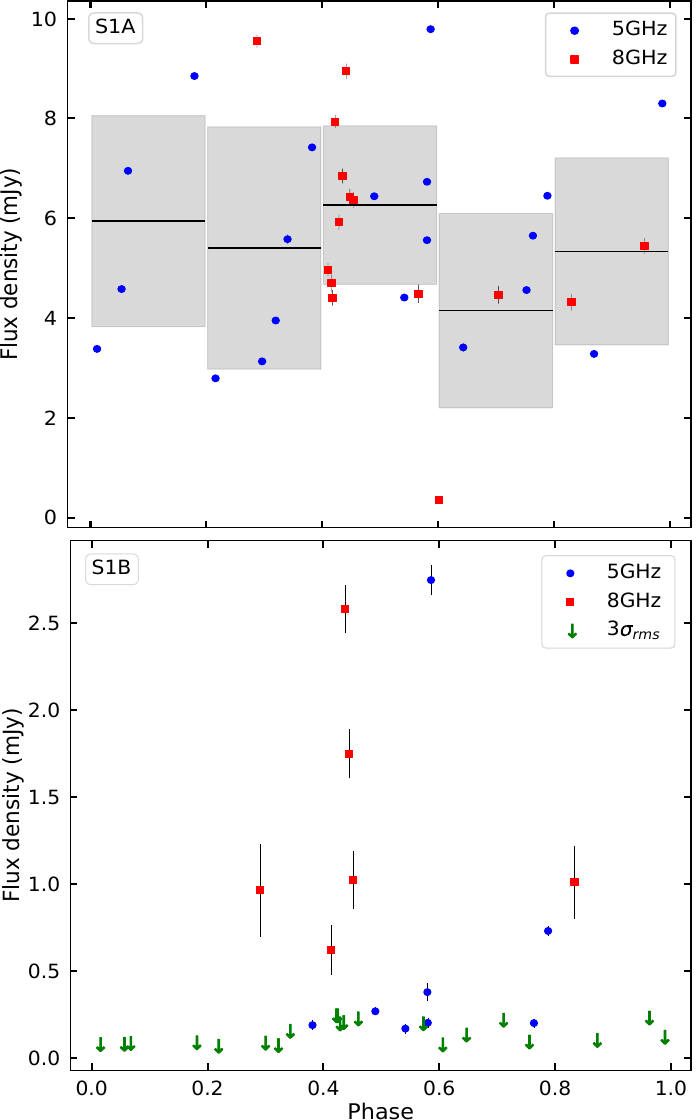}\\
\label{Flux_Phase}
\caption{Flux as a function of orbital phase for S1A (top) and S1B (bottom) at 5~GHz (blue dots) and 8~GHz (red squares). For S1A the bins (gray boxes) show the mean and standard deviation values of fluxes in widths of 0.2 in the orbital phase. For S1B the green arrows indicate the flux $3\,\sigma_{\rm rms}$ upper limits when the source is not detected. The error bars are shown and sometimes are smaller than the marker size.}
\end{figure}

Figure \ref{Flux_Phase} shows the flux of S1A and S1B as a function of the orbital phase, defined as the fractional part of the equation
$$\varphi=\frac{T-T_0}{P}$$
\noindent where $T_0$ is the time of periastron (JD 2457163.69), and P is the orbital period -- both taken from the fitting  results given in Table~\ref{T_A0}. 

\begin{figure}[!h]
\includegraphics[scale=0.42,trim=0 0 0 0,clip]{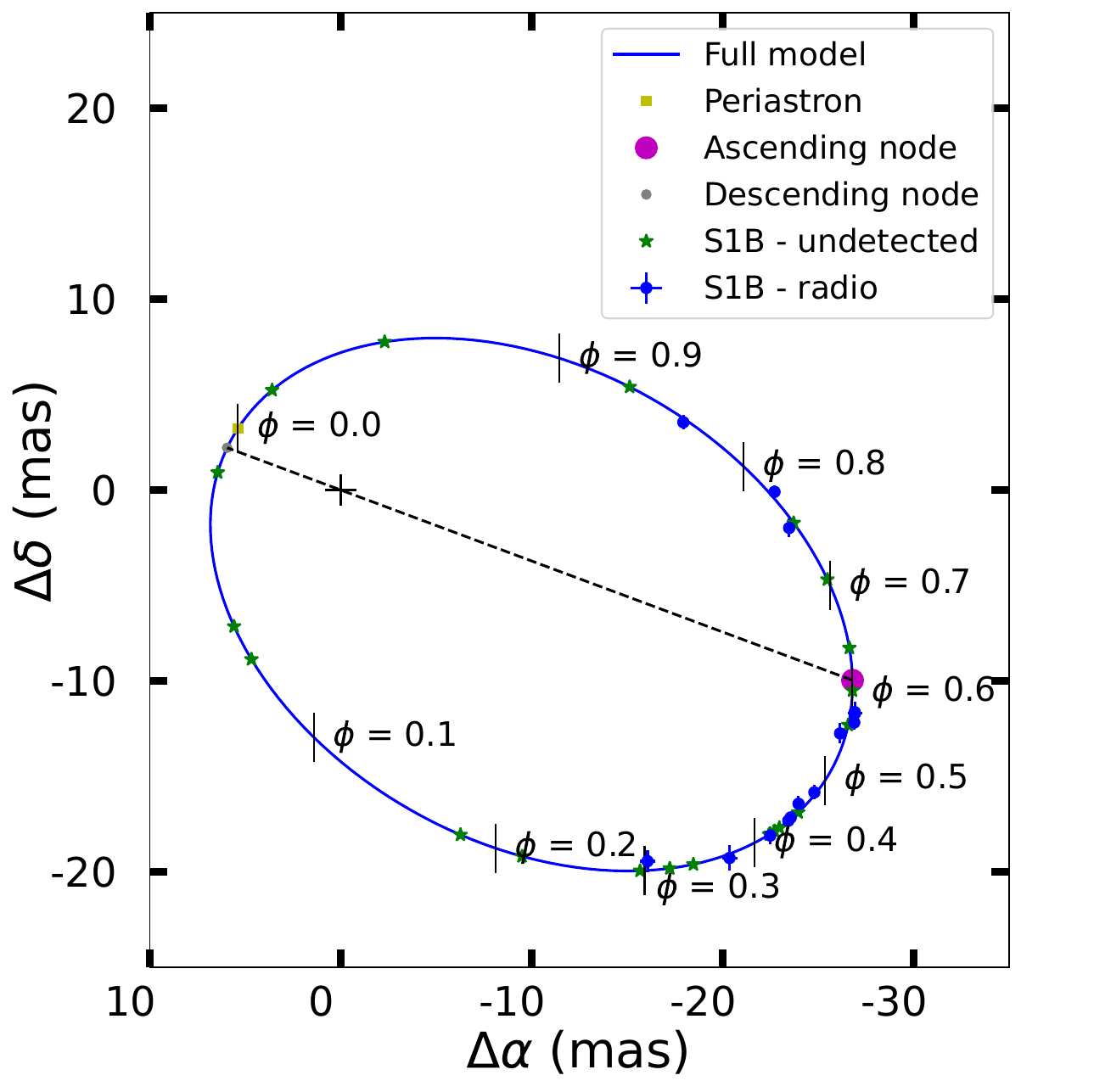}
\caption{Relative positions and orbital fit model. The green stars indicate the expected positions of S1B in undetected epochs. Orbital phases ($\phi$) are indicated with vertical black lines and labeled accordingly.} 
\label{S1or2}
\end{figure}

The projects BL128 and BT093 were recorded at a frequency of 8 GHz, while the GOBELINS project used 5 GHz, except for the first epoch which was observed at 8 GHz. As mentioned in Section \ref{sec:obs}, the DYNAMO--VLBA observations were also recorded at 5 GHz. To examine flux variations, we have to account for the fact that the data have been observed at two different frequencies. For S1A from the individual measured fluxes, we calculated the mean flux densities at both frequencies: $5.49\pm0.45$ mJy at 5 GHz and $6.95\pm0.57$ mJy at 8 GHz. This is roughly consistent with a flat spectrum over these frequencies, as also seen in previous VLA observations \citep{dzib2013}. Then, the flux values measured at 8 GHz were scaled to the 5 GHz values by adding the difference between the mean value at both frequencies. 

For S1A (top panel of Figure \ref{Flux_Phase}), there are no obvious variations with phase in the individual flux measurements. To search further for evidence of systematic flux variations, we averaged the data in bins of width 0.2 (in orbital phase). The horizontal black lines and vertical gray bars in the top panel of Figure \ref{Flux_Phase} indicate the mean and standard deviation in each bin, respectively. From this analysis, we do not find any evidence for variation with orbital phase. Variability at the level of about 50\% is present in the data, but it appears to be stochastic, with no relation to the orbital phase.

For S1B, on the other hand, detections (particularly associated with higher fluxes) appear to be clustered around an orbital phase value of 0.5 (bottom panel of Figure \ref{Flux_Phase}). More quantitatively, S1B was detected in 9 out of 14 observations (64\%) carried out in the range of orbital phases between 0.4 and 0.6, while it was detected in only 5 out of 21 observations (24\%) at orbital phases outside of this range. The lack of detections near periastron is illustrated further in figure \ref{S1or2} where detections are shown in blue and non-detections in green. The orbital phases ($\phi$) are indicated by tickmarks along the elliptical trajectory.

We can examine the significance of the detection rates in each orbital phase range by modeling the detectability of S1B with a binomial distribution. In this statistical model, if the events are independent and the probability of a success in a given experiment is $p$, the probability of $k$ successes in $n$ experiments is given by:

\[ p_{k,n} = {n \choose k} p^{k} (1-p)^{n-k}. \]

\noindent
S1B is detected in 14 out 35 experiments, so the overall probability of detection (a success) is $p = 0.4$. Given this probability and assuming that the chance of detection is independent of orbital phase, there is only a 4\% chance that 9 detections would be obtained out of 14 observations (the results obtained near apoastron), while there is only a 6\% chance that only 5 detections would be obtained out of 21 observations (the results obtained near periastron). We conclude that the higher detection fraction near apoastron than periastron is very unlikely to be the result of chance alone and that the source is intrinsically more likely to be detectable near apoastron.

\section{Discussion}

In this section, we mostly discuss the implications of our mass measurements for the properties of the S1 system. It is worthwhile mentioning at the outset that, to the best of our knowledge, no evidence for cicumstellar (or circumbinary) disks has ever been reported for this system.

\subsection{Spectral Energy Distribution of S1}
\label{SED0}

Before dynamical measurements became possible, the mass of S1 had been estimated from its photometry, and some spectroscopy, to be about 6~M$_\odot$. Let us briefly review these works. Based on early infrared observations, \citet{grasdalen1973} classified S1 as a late B star. It was later classified as B3V by \citet{elias1978} due to the surrounding H{\sc ii} region causing strong CO absorption in the 2.3~$\mu$m band. Combining spectroscopy and photometry, \citet{CohenKuhi1979} obtained a B2 classification. \citet{andre1988} estimated a B3.5 spectral type on the basis of the radio flux of the surrounding  H{\sc ii} region. From spectroscopic observations, \citet{bouvier1992} classified it as a B4 star since it exhibits strong H$_{\alpha}$ absorption but no lithium absorption; this same classification is adopted in \citet{andre1991}. Analysis by \citet{lada1984} of the spectral energy distribution implied that S1A has a luminosity of 1,500~$L_\odot$ and an effective temperature of 16,000~K consistent with a main-sequence star of spectral type B3-B5. Finally, spectroscopy by \citet{wilking2005} implied a spectral type earlier than B8 -- they adopted a B3 classification based on the effective temperature from \citet{lada1984}. The classification we just described implies a mass between 5 and 6 M$_\odot $ -- this is significantly higher than the value we derive dynamically. 

Recently, \citet{mookerjea2018} compiled photometric observations of S1A available through the astronomical catalog service VizieR to plot the SED of S1A and fit it with a reddened blackbody. They adopted a blackbody temperature of 17,000 K, which corresponds to the effective temperature of a B3/B4 ZAMS star, as suggested by the studies mentioned above. They assume a stellar radius $R = 4.35$ R$_\odot$ which, combined with the  temperature mentioned above, results in a luminosity $L = 1,420$ L$_\odot$. Their best fit implies a reddening $E(B-V) = 3.8$ mag, a selective extinction ratio $R_V \equiv A_V / E(B-V) = 3.35$, and a total $V$ band extinction $A_V = 12.7$ mag.

\begin{figure}[!hbt]
\includegraphics[width=0.5\textwidth]{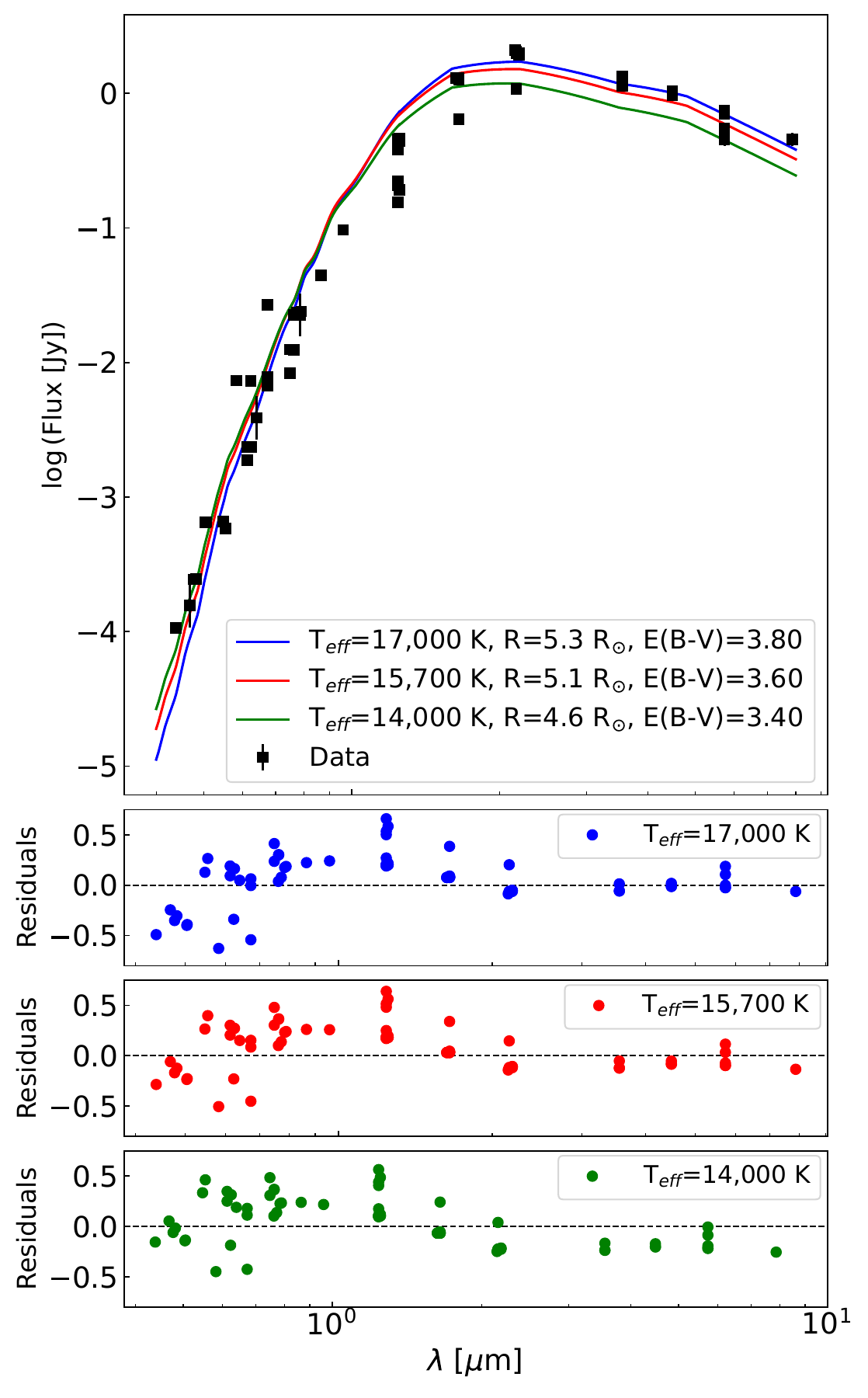}
\label{SED_S1}
\caption{Spectral energy distribution of S1A. The data points (black squares) are taken from  VizieR (see Table~\ref{vizier}) while the fits (color lines) are based on the extinction curves of \citet{rieke1989} and \citet{Fitzpatrick1999} [See text]. The figure also includes residual plots for each model (bottom panels).}
\end{figure}

In order to examine the compatibility between our dynamical mass of S1A and its observational properties, We performed an analysis similar to that of \citet{mookerjea2018}. We included the photometric points currently available in VizieR; which cover the optical and infrared bands from 0.4 $\mu$m to 8 $\mu$m. The complete list of included measurements is provided in table \ref{vizier} of Appendix~\ref{app:1}. We adopted a standard selective extinction ratio $R_V = 3.1$, and the extinction curves of \citet{Fitzpatrick1999} for the range between 0.4 $\mu$m and 1.25 $\mu$m and 
\citet{rieke1989} for the range between 1.25 $\mu$m and 8 $\mu$m. Figure \ref{SED_S1} shows that the data points can be reproduced by a range of effective temperatures, from about 14,000 K to 17,000 K. The specific values T$_{\text{eff}} = $ 14,000 K, 15,700, and 17,000 K shown in Figure \ref{SED_S1} were chosen because they correspond to spectral types B5, B4 and B3 according to \citet{Mamajek2013}. Values in the lower part of this range are consistent with the dynamical mass of S1 determined here. The extinction derived for the models shown in Figure \ref{SED_S1} vary from 10.5 (for $T_{\text{eff}} = 14,000$ K) to 11.8 ($T_{\text{eff}} = 17,000$ K). Plots of the residuals between the models and the photometric data are shown in the bottom panels.

\begin{figure}[!t]
\includegraphics[width=0.45\textwidth,clip]{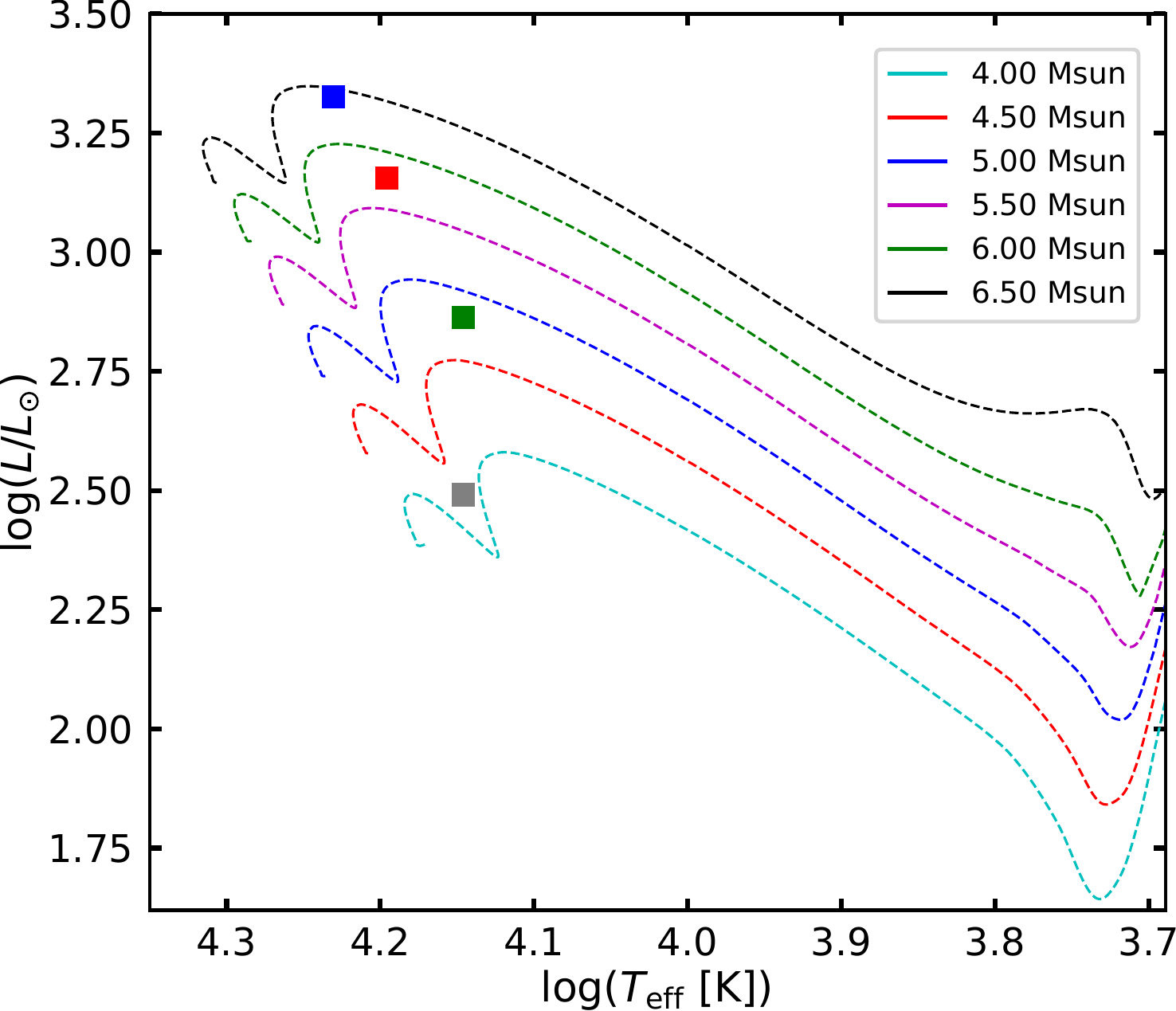}
\label{Pisa_s1}
\caption{Evolutionary tracks based on the PISA pre-main sequence stellar evolution models \citep{tognelli2011}. These models have a metalicity $Z=0.02$, helium abundance $Y=0.2880$, mixing-length $\alpha=1.90$ and deuterium abundance $X_D=4 \times 10^{-5}$. The blue, red, and green data squares indicate, respectively, stars with $T_{\text{eff}}$ = 17,000 K and $L$ = 2,120 L$_\odot$, $T_{\text{eff}}$ = 15,700  K and $L$ = 1,430 L$_\odot$, and $T_{\text{eff}}$ = 14,000 K and $L$ = 730 L$_\odot$. These correspond to the three SED models, of the same colors, in Figure \ref{SED_S1}. The grey square shows the locus of a star with $T_{\text{eff}}$ = 14,000 K, $L$ = 300 L$_\odot$, which would be consistent with a 4 $M_\odot$ star shortly before reaching the ZAMS.}
\end{figure}

\subsection{Stellar evolution models  with no rotation for S1A}
 
Dynamical masses obtained in binary systems are essential to constrain stellar evolution models. Nowadays, this is particularly important for young stars because pre-main sequence evolution models are less reliable and less mature than their main sequence counterparts \citep[e.g.,][]{hillenbrand2004,stassun2014}. Models for intermediate mass young stars are particularly uncertain given the paucity of existing observational constraints. It is, therefore, worthwhile to compare the observational properties of S1A with the predictions of theoretical models for the mass determined dynamically in the present paper. 

In Figure \ref{Pisa_s1}, we show evolutionary tracks for masses between 4 and 6 M$_\odot$, based on the PISA pre-main sequence stellar evolution models presented in \citet{tognelli2011}. The loci of S1A in the HR diagram for the three SED models shown in Figure \ref{SED_S1} are indicated in Figure \ref{Pisa_s1} as color points. 
 The luminosity and effective temperature of S1A appear to be consistent with masses between about 5 and 6 $M_\odot$, significantly higher than the measured dynamical mass of 4.1 $M_\odot$. An effective temperature of 14,000 K could be consistent with a mass of 4 $M_\odot$, but only provided that the luminosity was about 300 $L_\odot$ -- instead of the 730 $L_\odot$ obtained from modelling  SED. This could only be achieved if the radius of the star were of order 3 $R_\odot$ -- instead of 4.6 $R_\odot$ obtained from modelling SED.

For completeness, we have repeated the same analysis with the $Y^2$ (Yonsei-Yale) stellar models \citep{yi2001}, the Palla-Stahler models \citep{palla1999}, the PARSEC models \citep[][see below]{parsec2022A&A...665A.126N}, and the YaPSI YALE-POTSDAM stellar models \citep{spada2017}. We obtained similar conclusions to those reached with the PISA models: the theoretical predictions for a mass of 4 $M_\odot$ are inconsistent with the locus of S1A in the HR diagram. Specifically, the luminosity of S1A is at least twice larger than predicted by the models. Conversely, given the locus of S1A in the HR diagram, the theoretical models would predict a mass for S1A larger than measured dynamically, by 20 to 50\%. Interestingly, \citet{stassun2014} also reported systematic discrepancies between masses predicted by evolutionary models and measured in binary systems, but for stars with masses below 1 $M_\odot$. For masses above 1 $M_\odot$, the model predictions and observed masses were found to be in agreement at the level of about 10\%. We note, also, that previous dynamical mass measurements of intermediate-mass stars \citep[e.g.,][]{Johnston2019} did not find a significant discrepancy with evolutionary models.

To finish with this section, it is worth emphasizing that the location of S1A above and to the right of the main sequence on the HR diagram implies that it has not yet reached the main sequence. Since intermediate mass stars have a very short pre-main sequence lifetime, this places stringent constraints on the age of the S1 system. For instance, according to the evolutionary models of \citet{tognelli2011} shown in Figure \ref{Pisa_s1}, a 5 M$_\odot$ star reaches the main sequence in just about 1 Myr. The possible location of S1A along the track of a 5 M$_\odot$ star shown by the green square of Figure \ref{Pisa_s1} would indicate an age of 0.7 Myr according to the same models. Of course, we should proceed cautiously with obtaining age estimates from the models since we have just shown that the models cannot reproduce the location of S1A on the HR diagram given its measured dynamical mass. Nevertheless, it seems reasonable to conclude that the age of S1A is likely between 0.5 and 1 Myr.

\subsection{Stellar evolution models with rotation for S1A and magnetic activity}

The evolutionary models considered in the preceding section do not include rotation. To investigate the possibility that rotation might help resolve the discrepancy between the measured dynamical mass of S1A and the evolutionary tracks, we considered the PARSEC V2.0 evolutionary models presented by \citet{parsec2022A&A...665A.126N}. These models include recent re-calibrations of mixing by overshooting and rotation. They parametrize rotation via the $\omega$ parameter defined as the ratio between the stellar angular velocity $\Omega$ and the breakup angular velocity $\Omega_c$ ($\Omega_c$ is the angular velocity for which the centrifugal force becomes equal to the effective gravity at the equator). 

\begin{figure}[!t]
\includegraphics[width=0.45\textwidth,clip]{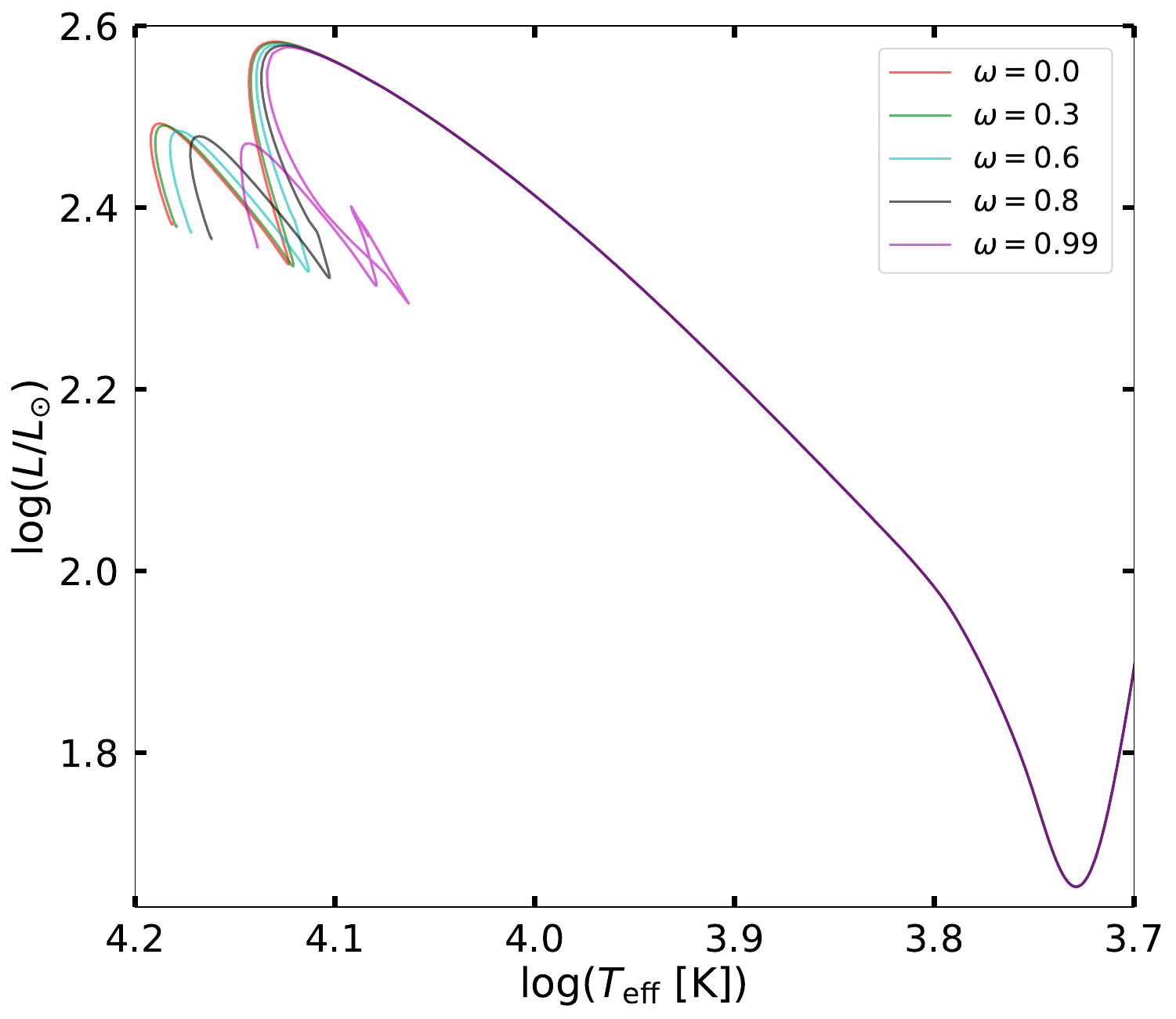}
\label{Parsec_s1}
\caption{PARSEC tracks \citep{parsec2022A&A...665A.126N} for a 4 M$_\odot$ Solar metallicity star with different rotation velocities from $\omega = 0$ to $\omega = 0.99$. See text for the definition of $\omega$.}
\end{figure}

Figure \ref{Parsec_s1} shows the PARSEC pre-main sequence tracks for a 4 M$_\odot$ star of Solar metallicity, appropriate for S1A, for $\omega$ values between zero (no rotation) and 0.99 (near breakup). As can be seen, the effect of rotation is imperceptible until the star approaches the ZAMS. There, the net effect of rotation is to make the star redder and slightly less luminous. Thus, rotation does not help reconcile the models with the dynamical mass of S1A since the models without rotation produced a luminosity that was already lower than expected for the dynamical mass of S1A.

The PARSEC models account for rotation but do not include magnetic fields and this may be an important missing element for the interpretation of S1A since the very existence of non-thermal radio emission requires the presence of a substantial superficial magnetic field (10$^2$--10$^3$ G; \citealt{andre1988}). The origin of this field is unclear. In low-mass stars, superficial magnetic fields of this magnitude are expected from dynamo theory, but this requires the presence of deep convective envelopes \citep{parker1955}. In intermediate mass stars such as S1A, however, the energy transport from the core to the surface is radiative, including during the pre-main sequence phase \citep{stellarevolution1990}. Several scenarios have been proposed to account for the presence of strong superficial magnetic fields in intermediate-mass stars. Perhaps the most accepted possibility is a fossil origin \citep{Schleicher+2023}. An alternative is the generation of magnetic field through the so-called Tayler-Spruit dynamo mechanism in a rotating star \citep{Devech2023}. Regardless of the origin of the magnetic field, its effect on the early evolution of intermediate-mass stars remains poorly investigated.

\subsection{The nature of S1B}

Our mass measurement and analysis of the SED of S1A implies that S1B is a young (age of order 0.5 to 1 Myr, assuming co-evality with S1A) Solar type star ($M \sim 0.8$ M$_\odot$). It is interesting to see if this is consistent with the few existing resolved infrared observations of S1A. The fluxes used in Figure \ref{SED_S1} and listed in Table \ref{tab:IR} correspond to a K band apparent magnitude of about 6.3 -- this is for the entire system, but we can also take this value for S1A alone since S1B is much weaker. \citet{richichi94} find a magnitude difference $\Delta K$ = 1.6 between S1A and S1B, while \citet{cheetham2015} report $\Delta K$ = 1.2 and 2.3 in their two separate observations. This results in an apparent K-band magnitude for S1B between 7.5 and 8.6. The distance modulus for S1 is $\mu = 5.7$ and the $A_V$ extinction of 12.7 reported from our fit to the SED of S1A corresponds to $A_K = 1.4$. Thus, we can estimate an absolute K band magnitude for S1B between 0.4 and 1.5.

According to the pre-main sequence models of \citet{Baraffe2015}, a 0.8 M$_\odot$ star with an age between 0.5 and 1 Myr should have an absolute K band magnitude between 1.5 and 2. We conclude that, within the errors, the observed K band magnitude of S1A is consistent with its dynamical mass. It is important to mention, in addition, that our age and magnitude estimates are quite uncertain. The measurements of \citet{richichi94} and \citet{cheetham2015} suggest significant variability (either intrinsic to S1B or caused by changing foreground absorption). The age of S1B is based on the assumption of co-evality of S1A and S1B and models for S1A that we have shown to be unable to reproduce the actual location of S1A in the HR diagram (Section 5.2). The extinction is assumed to be the same for S1A and S1B, but the presence of circumstellar material probably results in different absorptions for the two stars. Future dedicated infrared resolved observations ought to enable a more detailed comparison between the models and the properties of S1B.

\subsection{Comparison with Gaia Astrometry}

\begin{figure}[!t]
\includegraphics[scale=0.35]{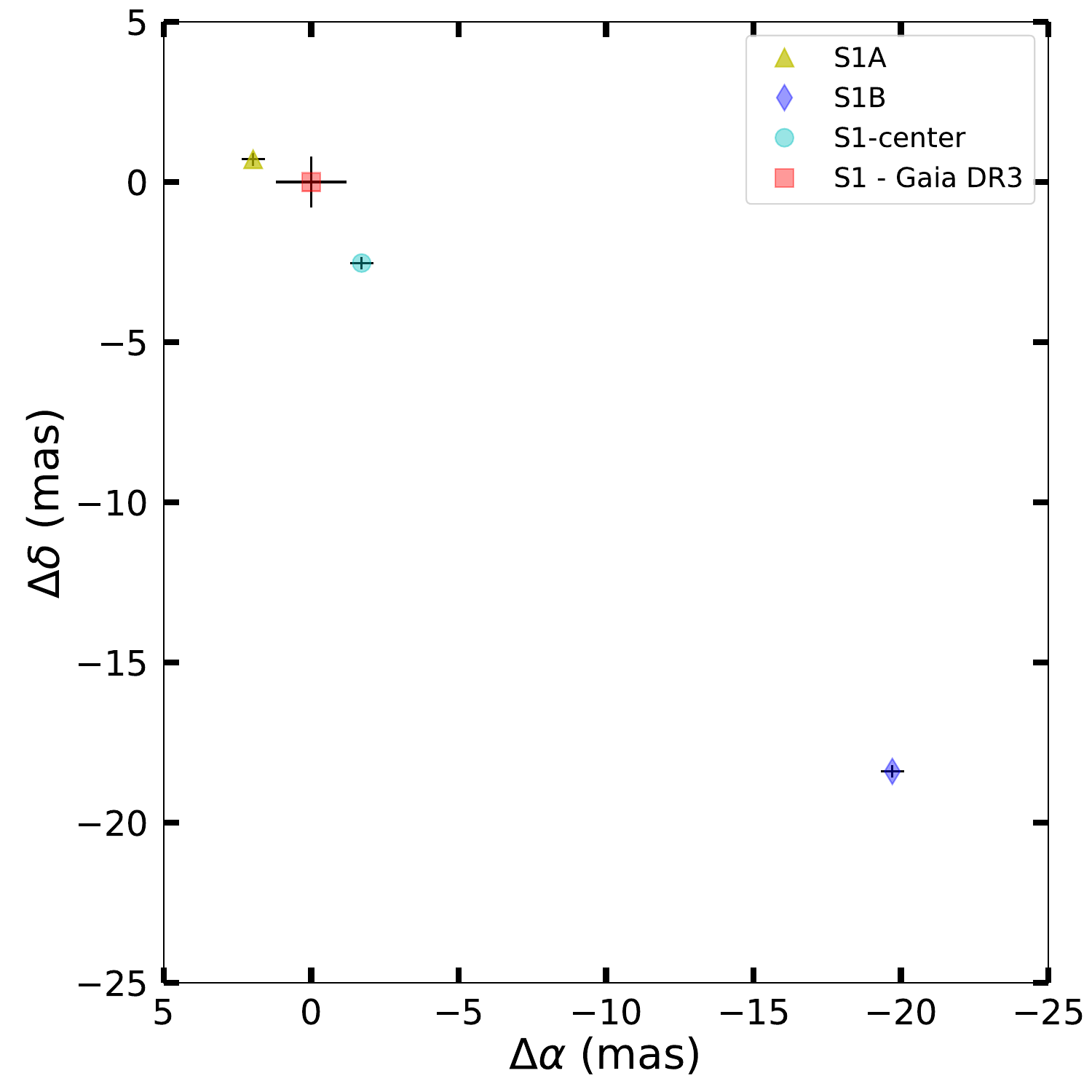}\\
\label{fig:gaia}
\caption{VLBA positions at epoch 2016.0 (the reference epoch  of the Gaia DR3 release) of S1A (yellow triangle), S1B (purple diamond), and the center of mass of the S1 system (cyan circle). The pink square indicates the Gaia DR3 position and is taken as the origin of the coordinate system. A black cross is shown associated with each position; the sizes of these black crosses are ten times the formal position errors in each case. }
\end{figure}

The Gaia satellite is the foremost astrometric mission at optical wavelengths \citep{gaia2016}. Its most recent data releases (Gaia DR3; \citealt{gaia2022} and Gaia EDR3; \citealt{gaia2021,lindegren2021}) were published in 2022 and 2021. The angular resolution of Gaia, $0\rlap{.}''18$ \citep{lindegren2021}, is insufficient to resolve the S1 system\footnote{It is estimated that the final Gaia resolution, in the upcoming fifth data release, will be $\sim0\rlap{.}''1$ (https://www.cosmos.esa.int/web/gaia/science-performance). It will, therefore, remain insufficient to spatially resolve the components in this system.} since the two components have a maximum separation of about 25\,mas (see Figure \ref{S1or}). The Gaia astrometric solution for S1 is listed in Table~\ref{tab:gaia} (Gaia source ID: Gaia DR3 6049161961233192576) where, for easy comparison, we also show  the astrometric results derived from our VLBA observations. The Gaia position does not coincide with the VLBA positions for S1A and S1B, as shown graphically in Figure \ref{fig:gaia}. It falls near the VLBA position for S1A, but is slightly displaced from it in the direction of S1B (or, equivalently, in the direction of the center of mass of the system). This is not unexpected: S1A is about 5 times more massive than S1B, and therefore presumably about at least 500 times more luminous. Thus, the photocenter of the S1 system traced by Gaia should indeed be located near S1A.

It is interesting to consider the consequences of the binarity of S1 on the Gaia results. The Gaia DR3 astrometric fitting assumes that S1 is a single star that moves in a linear and uniform fashion. Such a fit does not adequately describe the motion of S1 given the binary nature of the system. The Renormalised Unit Weight Error (RUWE) is a parameter provided by Gaia DR3 and used to assess if the single-star model provides a good fit to the astrometric observations \citep{Lindegren2018}. A RUWE value above 1.4 is considered an indication that a single-star model does not properly fit the astrometric data. In the case of  S1, the RUWE $=$ 2.75, confirming that a single-star model does not provide a satisfactory description of the data. The VLBA, on the other hand, resolves the components of the S1 system, and the astrometric fitting takes into account the orbital motion of the system. Thus, for S1, the VLBA astrometric results are certainly superior to those of Gaia. In addition, the VLBA covers a significantly longer time span (14.39 years against 2.76 for Gaia), further improving the quality of the VLBA results \citep[e.g., ][]{dzib2017}. The superiority of the VLBA astrometry is clearly seen in the accuracy of the parallax measurements (0.02 mas for the VLBA vs.\ 0.14 mas for Gaia). We also note that the two parallax values are barely consistent at the 3$\sigma$ level.

\begin{table*}[!th]
    \centering
        \caption{Comparison of DYNAMO-VLBA project astrometry with Gaia Astrometry}
    \begin{tabular}{c|ccc}
    \hline\hline
    Parameter           & Gaia DR3.    & VLBA & $\Delta$(VLBA -- Gaias EDR3)\\
    \hline
      $\alpha_{2016}$   & 246.642385343$^\circ\pm0.13\,$mas &  246.642384866$^\circ\pm0.04\,$mas& $-1.71\pm0.14$~mas\\ 
      $\delta_{2016}$   & $-24.391308440^\circ\pm0.08\,$mas &$-24.391309144^\circ\pm0.02\,$mas&$-2.53\pm0.08$~mas\\
VLBA $\alpha$ offset (S1A -- S1 center) &                             & 3.69 mas    & $\quad1.97\pm0.14$~mas\\
VLBA $\delta$ offset (S1A -- S1 center) &                             & 3.25 mas    & $\quad0.71\pm0.08$~mas\\
VLBA $\alpha$ offset (S1B -- S1 center) &                             & --17.99 mas & $-19.71\pm0.14$~mas\\
VLBA $\delta$ offset (S1B -- S1 center) &                             & --15.87 mas & $-18.40\pm0.08$~mas\\
      $\pi$             & $6.82\,\pm0.14\,$~mas& $7.28\,\pm0.02\,$~mas &    \\
      $\mu_\alpha$      & $-3.15\,\pm0.18\,$~mas yr$^{-1}$& $-2.536\,\pm0.003\,$~mas yr$^{-1}$ &  \\
      $\mu_\delta$      & $-26.68\,\pm0.15\,$~mas yr$^{-1}$& $-26.853\,\pm0.001\,$~mas yr$^{-1}$ &\\
      \hline
    \end{tabular}
    \label{tab:gaia}
\end{table*}

\subsection{Comparison with infrared astrometry}

The companion of S1A has also been inferred from infrared observations. The S1B component was first identified using the Lunar occultation (LO) method by \citet{richichi94}. Later, \citet{cheetham2015} also detected the companion using Sparse Aperture Masking (SAM). Both IR methods provide measurements of the total angular separation and the position angle (P.A.) between the components of the binary. The separations found in the IR observations are between 20 and 30 mas, consistent with the values we have measured at the VLBA (Figure~\ref{S1or}). From our best fit, we determined the orbital phase at the epochs of the IR observations, and the corresponding expected separations and P.A. The results are listed in Table~\ref{tab:IR} and shown graphically in Figure \ref{S1or}. We find that the IR observations were made close to apastron; within the errors, the expected separations and P.A. are in reasonable agreement with the IR results.

\begin{table*}[!th]
    \centering
        \caption{Comparison of Infrared and VLBA Astrometry}
    \begin{tabular}{ccccc|ccc}
    \hline\hline
    \multicolumn{5}{c}{Infrared}                 & \multicolumn{2}{|c}{VLBA}  &\\ 
 Obs. Date & JD & Method &Separation &   P.A.    &  Separation &     P.A.  & Orbital \\
 (dd.mm.yyyy) & &        & (mas)     & ($^\circ$)&    (mas)    & ($^\circ$)& phase\\
    \hline
13.06.1992& 2448786.5 & LO& $20\pm1$& 291$^a$ & $22.7\pm3.2$ &$268.6\pm7.2$&0.77\\
14.04.2012& 2456031.5 & SAM& $21.75\pm0.39$& $205.11\pm0.74$ &$21.0\pm3.1$ &$208.3\pm9.2$&0.21\\
09.03.2013& 2456360.5 & SAM& $28.89\pm3.47$& $268.54\pm2.13$ &$24.7\pm3.4$&$263.5\pm7.4$&0.73\\
      \hline
    \end{tabular}
    \label{tab:IR}\\
    Notes: $^a$ As pointed by \citet{richichi94}, we have used the correction of 180$^\circ$ to the scan angle reported by them. 
\end{table*}

\subsection{Flux variations}

In at least one young stellar system, V\,773~Tau, variations of the non-thermal radio flux synchronized with the orbital phase of the system have been well documented \citep{massi2002,torres2012}. In that specific case, the radio flux of both stars appears to increase substantially near the periastron of the system. In contrast, the radio emission of S1A shows no systematic variation with the orbital phase, while the flux of the low-mass companion S1B increases near the apastron (Section 4.2; Figure \ref{S1or2}).

The origin of this behavior of S1B is unclear. In the case of V\,773~Tau, it is plausible that the flux increase near periastron might be caused by increased flaring due to interactions between the individual magnetospheres when the stars approach each other \citep{massi2002}. Since flaring is thought to be initiated by magnetic reconnection events, it is indeed reasonable to expect an increase in the number of events when the two magnetospheres interact. To interpret the reverse effect observed here with the same class of model would require an unlikely scenario where reconnection events are more frequent when the magnetospheres are further apart. An alternative would be to invoke the presence of an optically thick region surrounding S1A with a radius of order 20 mas. Each time S1B plunged into this hypothetical region, its emission would be absorbed by the optically thick material explaining the lack of detection when the sources are closer than about 20 mas. The main difficulty with this scenario is to explain why this region does not absorb the emission from S1A itself. This requires a somewhat ad hoc morphology, with an inner hole, for the hypothetical absorbing structure --for instance a toroid seen close to face-on.

\section{Conclusions}

We have used seven new and 28 archival VLBA observations to study the S1 system in Ophiuchus,
and constrain the dynamical masses of its components. We detected the primary, S1A, in all 35 epochs, and the secondary, S1B, in 14 epochs. The best astrometric fit gives dynamical masses of M$_{\rm S1A}=4.11\pm0.10$~M$_\odot$ and M$_{\rm S1B}=0.83\pm0.01$~M$_\odot$. This is the most accurate mass estimation of S1 to date. The mass of S1A appears to be significantly smaller than previously suggested (of order 6 $M_\odot$). Reanalyzing the spectral energy distribution of S1, we conclude that it is compatible with effective temperatures in the range of 14,000 to 17,000~K. The location of S1 in the HR diagram suggests that it has not yet reached its zero-age main sequence, but we find that pre-main sequence evolutionary models are unable to explain its locus in the HR diagram for a mass of 4.1 M$_\odot$. For the models to pass through the allowed range of luminosity and effective temperature of S1A, a mass at least 20 to 50\% higher would be needed. Thus, evolutionary models appear to over-predict the mass of S1A by that amount.

A comparison between the VLBA radio astrometry and Gaia DR3 shows that the Gaia position falls close to, but is not coincident with the position of S1A measured by the VLBA. This is not unexpected since Gaia surely traces the motion of the photocenter of the system which should be located close to the primary. However, the fact that Gaia cannot resolve the system nor account for its orbital motion results in a parallax measurement that is significantly less accurate than, and barely compatible with, the VLBA result. Finally, we analyzed the variations of the flux density of S1A and S1B as a function of the orbital phase. We find that the flux of S1A remains stable over the entire orbit while  the flux of S1B appears to be higher when the system is close to apastron. The origin of this behavior is unclear.


\acknowledgements 
J.O., G.O.L., J.M.M. and L.F.R. acknowledge the financial support of CONACyT, M\'exico. L.L. acknowledges the support of DGAPA PAPIIT grants IN108324 and IN112820 and CONACyT-CF grant 263356.  S.A.D. acknowledge the M2FINDERS project from the European Research Council (ERC) under the European Union's Horizon 2020 research and innovation programme (grant No 101018682). P.A.B.G. acknowledges financial support from the São Paulo Research Foundation (FAPESP) under grants 2020/12518-8 and 2021/11778-9. The Long Baseline Observatory is a facility of the National Science Foundation operated under cooperative agreement by Associated Universities, Inc. The National Radio Astronomy Observatory is a facility of the National Science Foundation operated under cooperative agreement by Associated Universities, Inc. This work has made use of data from the European Space Agency (ESA) mission {\it Gaia} (\url{https://www.cosmos.esa.int/gaia}), processed by the {\it Gaia} Data Processing and Analysis Consortium (DPAC, \url{https://www.cosmos.esa.int/web/gaia/dpac/consortium}). Funding for the DPAC has been provided by national institutions, in particular the institutions participating in the {\it Gaia} Multilateral Agreement. This document was prepared using the collaborative tool Overleaf available at \url{https://www.overleaf.com/}. 

\facilities{VLBA \citep{napier1994}, Gaia \citep{gaia2016}} 
\software{AIPS \citep{Greisen2003}, Astropy\footnote{\url{https://www.astropy.org/}} \citep{astropy2013}, NumPy\footnote{\url{https://www.numpy.org/}} \citep{numpy2011}, SciPy\footnote{\url{https://www.scipy.org/}} \citep{scipy2014}, Matplotlib\footnote{\url{https://matplotlib.org/}} \citep{matplotlib2007}, and APLpy \citep{aplpy2012}}

\bibliography{references}

\begin{appendix}
\label{app:1}

\section{Orbital Parameters Analysis of S1 using MCMC}
\begin{figure*}[!hbt]
\includegraphics[width=\textwidth]{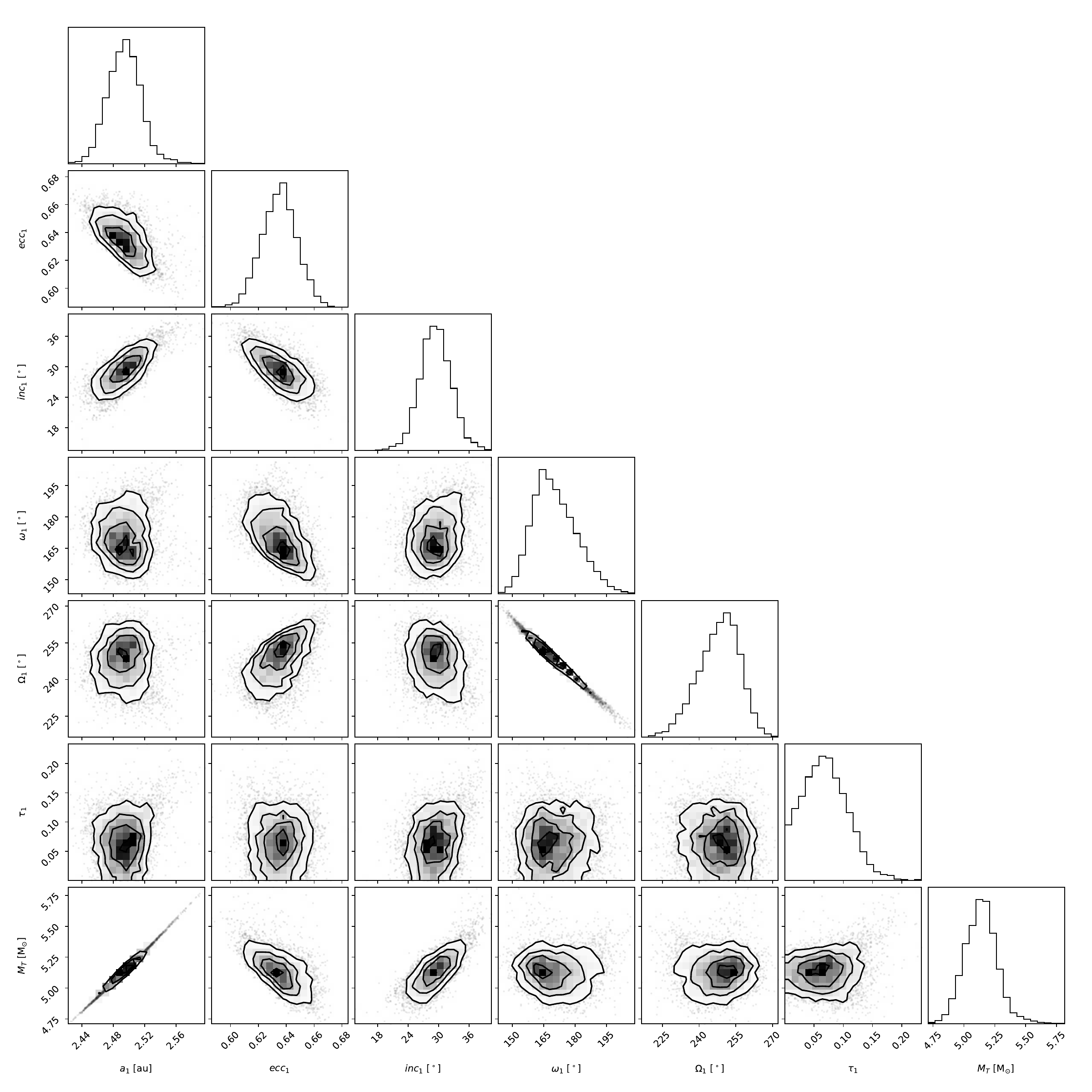}\\
\label{corner_plot}
\caption{Corner plot generated from the MCMC analysis using the Orbitize! library. The diagonal panels display 1D marginalized posterior distributions for each of the orbital parameters, including $a$, $e$, $i$, $\omega$, $\Omega$, $\tau$ and total mass. The off-diagonal panels illustrate 2D covariances between these parameters.}
\end{figure*}

As an additional step to rigorously quantify the key astrometric and orbital parameters, we implemented an additional approach that combines the power of two techniques: \textit{Orbits For The Impatient} (OFTI) and \textit{Markov Chain Monte Carlo} (MCMC). Both algorithms are integrated into the Orbitize! library\footnote{https://orbitize.info/en/latest/}, which is a specialized, open-source Python-based tool designed for characterizing orbits of binary astronomical systems and operates within a Bayesian analysis framework \citep{orbitize}.
For our analysis, we exclusively utilized 14 observations in which both components of the binary system were simultaneously detected, adhering to the necessary criteria for orbit analysis using Orbitize! OFTI works by generating a multitude of possible orbits based on the observed astrometric data. It effectively explores the parameter space, laying the groundwork for subsequent analysis with MCMC \citep{Blunt2017}. Based on the results of the OFTI fit, we defined priors for the orbital elements, including the semi-major axis ($a$), eccentricity ($e$), inclination ($i$), argument of periastron ($\omega$), position angle of nodes ($\Omega$), and epoch of periastron passage ($\tau$), which represents the fraction of the orbital period that has elapsed from a given reference epoch to periastron, within a range of [0, 1], and then proceeded to the MCMC phase.
To enhance the MCMC exploration, we employed a substantial ensemble of 10,000 walkers. These walkers independently traversed the parameter space, ensuring comprehensive sampling. To ensure the convergence of the MCMC chains, we executed a burn-in phase with 500 steps before extracting meaningful results. The results are presented in Table \ref{mcmc}, and the corner plot in Figure \ref{corner_plot}. This corner plot visually demonstrates the well-behaved nature of parameter sampling, further reinforcing the reliability of our results compared to MPFIT. We note, in particular that the total mass of the system derived from Orbitize! (5.14$\pm$0.13 M$_\odot$) is fully compatible with our MPIFIT result (4.94$\pm$0.11 M$_\odot$).
\begin{table}[!hbt]
	\renewcommand{\arraystretch}{1.1}
\caption{\footnotesize{Results obtained through the Orbitize! method.}}
	\label{mcmc}
	\centering
	\small
	\begin{tabular}{cccc} \hline \hline
		& Parameter & Value & Units \\ 
		\hline
		\multicolumn{2}{l}{Orbital Parameters}&&\\
		& $a$ & 2.494$\pm$0.022 & AU \\
           &P & 1.738$\pm 0.001 $ & years\\
           & $T_0$  & 2457164.29$\pm$0.02 &Julian date \\
            & $e$ & 0.64$\pm$0.01 & \\
            & $\Omega$&243.6$\pm$8.3& degrees \\
		& $i$ & 29.4$\pm$3.1& degrees \\
		& $\omega$ & 169.3$\pm$10.5& degrees \\
		& Total Mass &5.137$\pm$0.134 & $M_{\odot}$ \\
	      
	    \hline
	\end{tabular}  
	
\end{table}
\newpage

\begin{figure*}[!hbt]
\includegraphics[width=\textwidth]{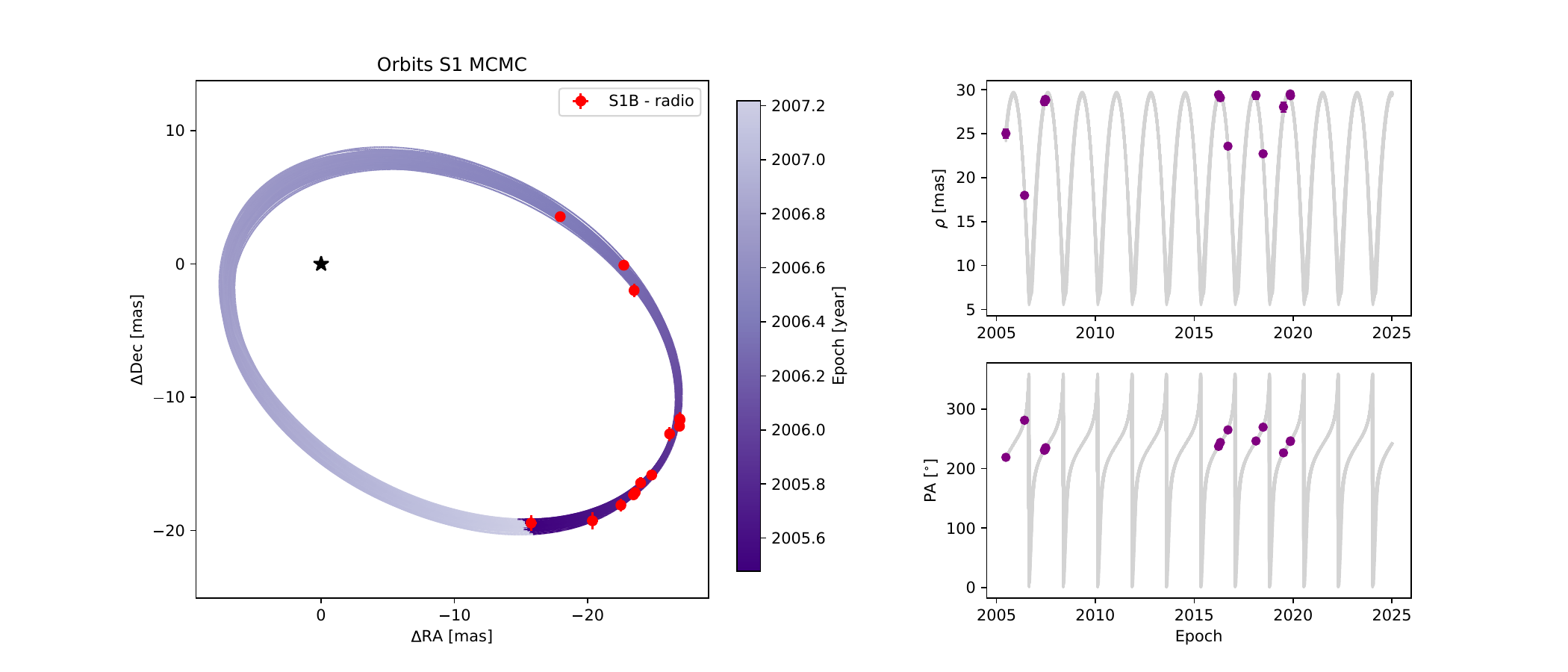}\\
\label{plot_mcmc}
\caption{Allowed orbital configurations for the binary system S1, derived from the MCMC analysis conducted using the Orbitize! library. The black marker (star) designates the position of the primary component, and the red points represent radio observations of the secondary component. The upper right panel depicts the evolution of the angular separation over time, while the lower right panel illustrates the variation in the position angle throughout the orbits.}
\end{figure*}

\newpage

\section{Photometric Data Table for SED Analysis}
\begin{center}
\footnotesize
\begin{longtable}{p{2cm} p{2cm} p{2.5cm} p{2.5cm} p{3.5cm}}
\caption{\footnotesize{VizieR S1 Photometric Data}} \\
\label{vizier}
\textbf{$\lambda$ ($\mu m$)} & {Flux (Jy)} & {Error flux (Jy)} & {SED Filter} & {ID Source} \\
\endfirsthead
\multicolumn{5}{c}%
{\tablename\ \thetable{} -- Continuation} \\
\textbf{$\lambda$ ($\mu m$)} & {Flux (Jy)} & {Error flux (Jy)} & {SED Filter} & {ID Source} \\
\endhead
\hline
2.19 & 1.99 & 0.06 & Johnson:K &\citet{referencia28}\\\hline
2.19 & 1.94 & 0.03 & Johnson:K &\citet{referencia9}\\
1.63 & 1.3 & 0.04 & Johnson:H &\\
1.25 & 0.444 & 0.004 & Johnson:J&\\\hline
2.16 & 1.98 & 0.03 & 2MASS:Ks&\citet{referencia12} \\
1.65 & 1.28 & 0.04 & 2MASS:H &\\
1.24 & 0.456 & 0.008 & 2MASS:J& \\
7.87 & 0.455 & 0.05 & Spitzer/IRAC:8.0 &\\
5.73 & 0.741 & 0.063 & Spitzer/IRAC:5.8& \\
4.49 & 1.04 & 0.06 & Spitzer/IRAC:4.5& \\
3.55 & 1.34 & 0.07 & Spitzer/IRAC:3.6& \\\hline
0.789 & 0.0239 & \nodata & Cousins:I &\citet{referencia21}\\\hline
 1.25 & 0.191 &0.003 &UKIDSS:J&\citet{referencia11} \\\hline
 2.16 & 2 & 0.04 & 2MASS:Ks &\citet{referencia29}\\
 1.65 & 1.31 & 0.04 & 2MASS:H &\\
 1.24 & 0.451 & 0.008 & 2MASS:J& \\
 5.73 & 0.701 & 0.02 & Spitzer/IRAC:5.8 &\\
 4.49 & 0.969 & 0.018 & Spitzer/IRAC:4.5& \\
 3.55 & 1.14 & 0.03 & Spitzer/IRAC:3.6& \\\hline
 0.784 & 0.0226 & 0.0083 & POSS-II:i&\citet{referencia2} \\
 0.64 & 0.00388 & 0.00146 & POSS-II:F& \\
 0.468 & 0.000156 & 5.9E-05 & POSS-II:J& \\\hline
 2.16 & 2.04 & 0.04 & 2MASS:Ks &\citet{referencia7}\\
 1.65 & 1.31 & 0.04 & 2MASS:H &\\
 1.24 & 0.442 & 0.004 & 2MASS:J& \\
 0.763 & 0.0125 & 0.0002 & SDSS:i &\\
 0.673 & 0.00668 & 1E-05 & Gaia:G &\\
0.673 & 0.0269 & 0.0001 & Gaia:G &\\ 
 0.625 & 0.00236 & 3E-05 & SDSS:r& \\
 0.482 & 0.000245 & 5E-06 & SDSS:g &\\\hline

1.65 & 1.31 & 0.04 & 2MASS:H&\citet{referencia10}\\
1.24 & 0.451 & 0.008 & 2MASS:J &\\\hline
5.73 & 0.455 & 0.05 & Spitzer/IRAC:5.8 &\citet{referencia30}\\
5.73 & 0.741 & 0.063 & Spitzer/IRAC:5.8 &\\
4.49 & 1.04 & 0.06 & Spitzer/IRAC:4.5& \\
3.55 & 1.34 & 0.07 & Spitzer/IRAC:3.6& \\\hline
2.15 & 2.1 & \nodata & MKO:Ks& \citet{referencia23}\\\hline
0.554 & 0.000582 &\nodata & Johnson:V&\citet{referencia3} \\\hline
2.16&	2.04&	0.05&2MASS:Ks&\citet{referencia8} \\
1.65&	1.27&	0.07&2MASS:H	&\\
1.24&	0.155&	\nodata&2MASS:J	&\\
1.24&	0.206&	\nodata&2MASS:J	&\\
1.24&	0.442&	0.008&2MASS:J	&\\
1.24&	0.223&	\nodata	&2MASS:J &\\\hline

  0.547 & 0.000656 & \nodata &UVOT:V&\citet{referencia14} \\
  0.439 & 0.000106 &  \nodata&UVOT:B& \\\hline
  
 0.763 & 0.0124 & 0.001 & SDSS:i' &\citet{referencia13} \\\hline
  0.96 & 0.0968 &\nodata & PAN-STARRS/PS1:y&\citet{referencia15} \\
  0.865 & 0.0445 &\nodata & PAN-STARRS/PS1:z& \\
 0.748 & 0.0125 & 0.0002 & PAN-STARRS/PS1:i& \\
 0.613 & 0.00235 & 3E-05 & PAN-STARRS/PS1:r& \\
 0.477 & 0.000244 & 5E-06 & PAN-STARRS/PS1:g& \\\hline
  0.673 & 0.00779 & 1E-05 & Gaia:G&\citet{referencia4} \\\hline
 0.748 & 0.00833 & 5E-05 & PAN-STARRS/PS1:i &\citet{referencia26}\\
 0.613 & 0.00188 & 2E-05 & PAN-STARRS/PS1:r &\\\hline

 0.772 & 0.0234 & 0.0001 & GAIA/GAIA2:Grp&\citet{referencia5} \\
 0.623 & 0.0073 & 4E-06 & GAIA/GAIA2:G& \\
 0.505 & 0.000654 & 3E-06 & GAIA/GAIA2:Gbp& \\\hline

2.16 & 2.01 & 0.03 & 2MASS:Ks &\citet{referencia17}\\
1.65 & 1.31 & 0.04 & 2MASS:H&\\
1.24 & 0.451 & 0.008 & 2MASS:J& \\\hline
2.16 & 2.01 & 0.03 & 2MASS:Ks &\citet{referencia18}\\
 1.63 & 1.3 & 0.04 & Johnson:H &\\
 1.25 & 0.46 & 0.008 & Johnson:J &\\\hline
 0.762 & 0.0226 & 0.0001 & GAIA/GAIA3:Grp &\citet{referencia6}\\
 0.582&	0.00732	&2.00E-05&GAIA/GAIA3:G\\
 0.504 & 0.000647 & 3E-06 & GAIA/GAIA3:Gbp& \\\hline
  5.73 & 0.548 & 1.6E-04 & Spitzer/IRAC:5.8&\citet{referencia16} \\\hline
 2.16 & 2.01 & 0.03 & 2MASS:Ks &\citet{referencia19}\\
 1.63 & 1.3 & 0.04 & Johnson:H &\\
 1.25 & 0.46 & 0.008 & Johnson:J& \\\hline
 2.19 & 1.94 & 0.02 & Johnson:K &\citet{referencia20}\\
 1.63 & 1.3 & 0.04 & Johnson:H &\\
 1.25 & 0.46 & 0.005 & Johnson:J &\\\hline

 2.16 & 1.08 & 0.05 & 2MASS:Ks&\citet{referencia34}\\
 1.65 & 0.644 & 0.03 & 2MASS:H &\\
 1.24 & 0.38 & 0.017 & 2MASS:J &\\\hline
\end{longtable}

\end{center}

\end{appendix}

\end{document}